\documentclass[aps,prx,reprint,superscriptaddress,floatfix,longbibliography]{revtex4-2}

\usepackage{amsmath,amssymb,graphicx,bm}
\graphicspath{{./}{Plots/}{Plots/Appendix/}{Plots/Stability/}}
\usepackage[colorlinks=true,linkcolor=blue,citecolor=blue,urlcolor=blue]{hyperref}
\usepackage[dvipsnames]{xcolor}
\usepackage{booktabs}
\usepackage{wrapfig}
\usepackage{needspace}
\usepackage{multirow}
\usepackage{braket}
\usepackage{soul}

\newcommand{\rv}{r_{\rm v}}
\newcommand{\ro}{r_{\rm o}}

\newcommand{\Nsurf}{N_{\rm surf}}

\begin{document}

\title{Stitched surface codes: planar concatenation through GHZ corridors}

\author{Mohamed A.\ Shalby}
\email{mshal007@ucr.edu}
\affiliation{Google Quantum AI, Santa Barbara, CA 93117, USA}
\affiliation{Department of Physics and Astronomy, University of California, Riverside, CA 92507, USA}
\author{Alexandre Bourassa}
\affiliation{Google Quantum AI, Santa Barbara, CA 93117, USA}
\author{Amir Karamlou}
\affiliation{Google Quantum AI, Santa Barbara, CA 93117, USA}
\author{Leigh Martin}
\affiliation{Google Quantum AI, Santa Barbara, CA 93117, USA}
\author{Alexis Morvan}
\affiliation{Google Quantum AI, Santa Barbara, CA 93117, USA}
\author{Nicholas Noll}
\affiliation{Google Quantum AI, Santa Barbara, CA 93117, USA}
\author{Craig Gidney}
\affiliation{Google Quantum AI, Santa Barbara, CA 93117, USA}
\author{Leonid P.\ Pryadko}
\affiliation{Google Quantum AI, Santa Barbara, CA 93117, USA}

\date{\today}

\begin{abstract}
We present a fault-tolerant quantum-memory architecture that concatenates four distance-$t$ rotated surface-code patches with the $[\,\![4,2,2]\,\!]$ outer code, encoding two logical qubits at code distance $d=2t$ in a physical footprint of 
$N<2(2t+1)^2$ qubits on a square lattice with nearest-neighbor connectivity.
The outer-code stabilizer generators $\overline{XXXX}$ and $\overline{ZZZZ}$, each of physical weight $4t$, are measured \emph{in situ} via verified Greenberger--Horne--Zeilinger (GHZ) ancillae distributed through narrow corridors between adjacent patches, without lattice surgery or long-range gates.
The outer-code gadgets introduce hyperedge fault mechanisms, but by design each decomposes into graphlike components, so the complete syndrome is decoded by a single minimum-weight-perfect-matching pass.
Circuit-level simulations show logical error rates that scale as $\sim p^{\,t}$, consistent with distance-$2t$ protection.
Finally, we implement the construction experimentally on a 105-qubit Willow superconducting processor in both the $ZZ$ and $XX$ memory bases.  With the correlated-matching decoder, increasing the code distance from $d=4$ to $d=6$ reduces the block logical error rate by a factor of $1.52$. The construction thus provides a near-term testbed for concatenated architectures beyond the standard surface code.
\end{abstract}

\maketitle

\section{Introduction}
\label{sec:intro}

Rotated surface codes are the leading quantum error-correcting code (QECC) family for superconducting hardware, owing to their two-dimensional nearest-neighbor (2D nn) connectivity on a square lattice and high thresholds~\cite{Fowler2012PRA,Dennis2002,Kitaev2003}.
In the standard architecture, each patch of distance~$d$ encodes a single logical qubit, $k=1$, in $n_{\rm surf}=d^2$ data qubits using $d^2-1$ ancillae, for a total of $\Nsurf = 2d^2 - 1$ physical qubits. Stabilizer generators are measured by depth-four circuits (counting two-qubit gates only), and the logical Clifford group is realized, e.g., by lattice surgery between adjacent patches~\cite{Horsman2012,FowlerGidney2018,LitinskiVonOppen2018,Litinski2019}.
Single-patch surface-code memory has now been realized below threshold on superconducting hardware, with exponential suppression up to distance seven~\cite{Google2023Scaling,Google2025BelowThreshold} (also see earlier distance-three demonstrations on independent platforms~\cite{Krinner2022,Zhao2022} and surface-code logical qubits on neutral-atom arrays~\cite{Bluvstein2024}).

Beyond surface codes, two large families of fault-tolerant QECCs are known.
First, quantum low-density parity-check (qLDPC) codes sacrifice the locality in 2D but may encode $k>1$ logical qubits per physical block, in particular, they may have finite asymptotic rates $k/n$.  For fault tolerance, it is sufficient that the distance scales as a logarithm of $n$ or higher~\cite{KovalevPryadko2013,DumerKovalevPryadko2015}; many families of such codes are known~\cite{FreedmanMeyerLuo2002,BreuckmannTerhal2016,Breuckmann2017SemiHyperbolic,TillichZemor2014}, including asymptotically good families with finite relative distances $d/n$~\cite{PanteleevKalachev2022,LeverrierZemor2022,DinurHsiehLinVidick2023}.  Second, concatenated codes, historically the first route to fault tolerance~\cite{Shor1996,KnillLaflamme1996,AharonovBenOr1999}, encode each qubit of an outer code as a logical qubit of an inner code, and also can attain asymptotically constant overhead in a fault-tolerant setting~\cite{Yoshida2025}.  Neither construction suits hardware with limited, fixed qubit connectivity, such as superconducting qubits.  In particular, bivariate bicycle codes~\cite{Bravyi2024BB} require  bilayer connectivity graphs with at least two non-local edges per qubit node. 
\begin{figure}[tbp]
  \centering
  \includegraphics[width=0.95\linewidth]{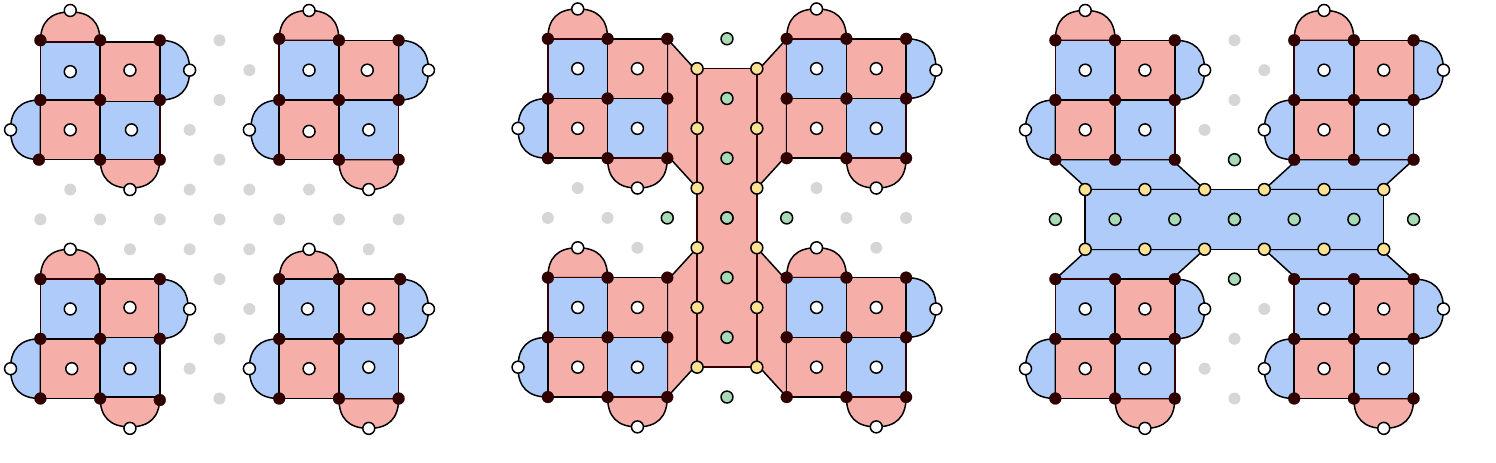}
  \caption{Stabilizer structure for the $t=3$ stitched code, with four rotated surface-code patches connected by the vertical and horizontal GHZ corridors for the outer-code generators. Filled and open circles, respectively, represent data and ancillary qubits.}
  \label{fig:layout_d3}
\end{figure}

A hybrid line of recent work pairs inner surface-code patches with outer codes to raise the encoding rate without abandoning the inner code's hardware-friendly structure.
Yoked surface codes~\cite{Gidney2025Yoked} measure the stabilizer generators of a high-rate outer code via lattice surgery to ancillary ``workspace'' patches; the qubit savings (nearly $3\times$ in their reported regime) scale with the outer-block size and only materialize once the outer dimension is large enough to amortize the workspace overhead.
Similarly, hierarchical memories~\cite{Pattison2025} concatenate qLDPC outer codes with surface-code inner blocks via surgery-style measurements.
Unfortunately, these constructions, designed for the large-scale fault-tolerant regime, require too many physical qubits to demonstrate non-trivial distance scaling and/or an advantage over surface codes on existing or near-term superconducting devices~\cite{Gidney2025Yoked,Pattison2025}.  

The goal of this work is to demonstrate a similar concatenated construction (see Fig.~\ref{fig:layout_d3}), both theoretically and experimentally, but suitable for existing devices.
We concatenate four distance-$t$ rotated surface-code patches with the smallest non-trivial error-detecting CSS code, $[\,\![4,2,2]\,\!]$~\cite{Vaidman1996,GrasslBethPellizzari1997,Knill2005}, obtaining a $[\,\![4t^2,2,2t]\,\!]$ code of distance $d=2t$ that encodes $k = 2$ logical qubits in a footprint of $N = 2d^2 + 4d - 3$ physical qubits on a square lattice with nn connectivity, see Fig.~\ref{fig:layout_d3}.
The two outer-code stabilizer generators $G_X=\overline{XXXX}$ and $G_Z=\overline{ZZZZ}$, each of  weight $4t$, are measured sequentially by verified GHZ resource states prepared on narrow \emph{corridors} of ancillary qubits between adjacent patches.
The corridors thread between patches using only nearest-neighbor gates and are operated alongside the standard surface-code stabilizer cycles, so the inner patches' bulk stabilizer groups remain intact throughout outer-code extraction. The corridor overhead, $8t+1$ qubits, grows only linearly with the patch distance $t$.

The GHZ corridors introduce a potential structural complication for decoding.
A single circuit-level fault inside the corridor can simultaneously trigger detectors in the corridor itself and in two or more adjacent patches, producing hyperedge fault mechanisms in the detector error model~\cite{Gidney2021Stim}.
Decoders that natively handle hyperedges---e.g., belief propagation with ordered statistics (BP+OSD)~\cite{PanteleevKalachev2021,Roffe2020} and search-based decoders such as Tesseract~\cite{Tesseract2025}---can handle these mechanisms more accurately than minimum-weight perfect matching (MWPM), but at substantially greater runtime, which limits their use at scale and at real-time hardware cycle times.
Here we construct the corridor checks and verifiers to ensure that every such mechanism decomposes into graphlike components, so that the complete syndrome, including the outer-code detectors, is decoded by a single MWPM pass~\cite{Higgott2025SparseBlossom}, optionally with correlated reweighting~\cite{Fowler2013Correlated}.

The remainder of the paper is organized as follows.
Section~\ref{sec:prelim} reviews the relevant background: surface-code preliminaries, the concatenation framework, the $[\,\![4,2,2]\,\!]$ outer code, and standard schemes for outer-stabilizer measurement.
Section~\ref{sec:arch} specifies our architecture: the four-patch layout, the GHZ corridor preparation and verification protocol, and the scheduling that interleaves outer-code rounds with inner-patch cycles.
Section~\ref{sec:decoding} describes how the complete syndrome is decoded by matching.
Section~\ref{sec:results} presents circuit-level simulation results under the SI1000 error model~\cite{GidneyNewmanFowlerBroughton2021}, including the verification-depth tradeoff, distance scaling, and footprint comparisons.
Section~\ref{sec:experimental} reports a hardware implementation on a 105-qubit superconducting Willow processor in two memory bases.  Specifically, we implement $d=4$ and $d=6$ stitched codes, and demonstrate robust below-threshold scaling of the logical error rates.
Section~\ref{sec:outlook} closes with the discussion of geometric and modular advantages, experimental considerations, and open problems.  Additional theoretical and experimental details are presented in the Appendices.

\section{Preliminaries}
\label{sec:prelim}

\subsection{Rotated surface codes}
\label{sec:prelim_surface}

The rotated surface code of distance $t$ encodes one logical qubit in $n = t^2$ data qubits, or $\Nsurf = 2t^2 - 1$ physical qubits including syndrome ancillae, arranged on a square lattice with open boundaries~\cite{BombinMartinDelgado2007,Horsman2012,Litinski2019}.
The code is a CSS code whose stabilizer generators are weight-four $X$-type plaquettes and weight-four $Z$-type plaquettes in the bulk, together with weight-two operators on opposite pairs of boundaries, see Fig.~\ref{fig:layout_d3} (left).
Minimal-weight logical operators $\overline{X}$ and $\overline{Z}$ are strings of weight $t$ connecting the two pairs of boundaries.
A single round of syndrome extraction interleaves the $X$- and $Z$-type measurements in a depth-four circuit (depth-six if basis-change layers are counted) between data and adjacent syndrome qubits~\cite{Fowler2012PRA,Dennis2002}.

\subsection{Concatenated quantum codes}
\label{sec:prelim_concat}

The concatenation of an outer code $\mathcal{Q}_{\rm o} = [\,\![n_{\rm o},k_{\rm o},d_{\rm o}]\,\!]$ with an inner code $\mathcal{Q}_{\rm i} = [\,\![n_{\rm i},k_{\rm i},d_{\rm i}]\,\!]$ encodes each qubit of $\mathcal{Q}_{\rm o}$ as a logical qubit of $\mathcal{Q}_{\rm i}$~\cite{KnillLaflamme1996,Knill2005}, which gives a code $\mathcal{Q}_{\rm o} \star \mathcal{Q}_{\rm i}$ with parameters
\begin{equation}
[\,\![n_{\rm o} n_{\rm i},\; k_{\rm o} k_{\rm i},\; d]\,\!],\qquad d \geq d_{\rm o}\, d_{\rm i}.
\label{eq:distance-product}
\end{equation}
The distance-product bound is due to the fact that an outer logical operator must have support on at least $d_{\rm o}$ inner blocks, and each nontrivial inner logical representative has weight at least $d_{\rm i}$. For stochastic circuit-level noise this code-distance statement translates into logical-failure scaling governed by the minimum number of physical faults needed to realize such a representative.

\subsection{The outer code and full code construction}
\label{sec:prelim_422}

The outer $[\,\![4,2,2]\,\!]$ code is the smallest CSS code with a non-trivial distance $d_{\rm o}=2$; the code \emph{detects} any single-qubit error but may not be able to correct it.
There are two stabilizer generators, 
\begin{equation}
G_X = X_1 X_2 X_3 X_4,\qquad G_Z = Z_1 Z_2 Z_3 Z_4,
\label{eq:422-stabs}
\end{equation}
which gives $k_{\rm o} = 2$ logical qubits. The logical operators can be chosen as
\begin{align}
\overline{X}_1 &= X_1 X_2,\quad \overline{X}_2 = X_1 X_3,\label{eq:422-logicals-X}\\
\overline{Z}_1 &= Z_1 Z_3,\quad \overline{Z}_2 = Z_1 Z_2.
\label{eq:422-logicals}
\end{align}

While not used in this work, we notice that logical Clifford gates on the encoded pair are accessible through transversal CNOT between two $[\,\![4,2,2]\,\!]$ blocks, a transversal Hadamard ($H^{\otimes 4}$, which performs $\overline{H}_1 \otimes \overline{H}_2$ composed with a swap of the two encoded qubits), physical-qubit permutations that realize $\overline{\rm CNOT}_{12}$ within a single block, and the transversal phase gate $S\otimes S^\dagger\otimes S^\dagger\otimes S$, which realizes $\overline{\rm CZ}_{12}$~\cite{Gottesman2016Small}; this transversal-Clifford structure makes the $[\,\![4,2,2]\,\!]$ code a standard first level of concatenation in fault-tolerance demonstrations~\cite{Knill2005}.

For the inner code we use the rotated surface code with parameters  $[\,\![t^2,1,t]\,\!]$.
The concatenated stabilizer code has length $n=4t^2$, dimension $k=2$, and distance $d=2t$ which saturates the bound in Eq.~\eqref{eq:distance-product}.
The logical operators are obtained by composing the outer-code logicals of Eqs.~\eqref{eq:422-logicals-X} and~\eqref{eq:422-logicals} with the inner-code logical strings of weight $t$: e.g.\ $\overline{\overline{X}}_1 = \overline{X}^{(1)} \otimes \overline{X}^{(2)}$, where $\overline{X}^{(j)}$ acts on the inner surface-code patch $j$. The minimum-weight logical of the concatenated code therefore has weight $2t$, which gives the code distance.

\subsection{Outer-stabilizer measurement}
\label{sec:prelim_measurement}

Three standard schemes measure a weight-$w$ stabilizer generator fault-tolerantly with the help of ancillary qubits: Shor-style cat ancilla~\cite{Shor1996,AliferisGottesmanPreskill2008}, Steane-style encoded ancilla~\cite{Steane1997,KnillLaflamme1996}, and Knill-style ancilla teleportation~\cite{Knill2005}.
All three reduce the syndrome-extraction circuit to transversal CNOT gates between an ancilla register and the data qubits supporting the operator being measured.
Steane and Knill schemes scale ancilla overhead with the size of an encoded block and are well-suited when the data block is itself encoded. Shor-style cat ancillae use a single GHZ-type resource state and incur the lowest overhead.
For weight-$w$ stabilizer measurement, a Shor-style scheme prepares a $w$-qubit GHZ state
\begin{equation}
\ket{\mathrm{GHZ}_w} = \tfrac{1}{\sqrt{2}}\bigl(\ket{0}^{\otimes w} + \ket{1}^{\otimes w}\bigr),
\end{equation}
applies transversal CNOTs from ancilla to data (for an $X$-type measurement; CNOTs from data to ancilla for $Z$-type), and measures the ancilla in the $X$ basis to read out the joint $X^{\otimes w}$ eigenvalue.
The dangerous fault channel is an $X$ error on the GHZ ancilla, which propagates into the data through the transversal CNOT and can grow into a weight-2 or higher error on the supported data qubits (a \emph{hook} error~\cite{Dennis2002}).
$Z$ errors on the ancilla commute with the subsequent CNOTs and merely flip the readout, behaving as classical measurement errors.

Suppressing the dangerous $X$ errors is accomplished by adding $\rv$ rounds of $ZZ$-parity \emph{verification} between adjacent GHZ qubits before the transversal CNOTs into data~\cite{DiVincenzoAliferis2007}.
Unitary preparation of large GHZ states on superconducting processors has reached 60 and 120 qubits~\cite{Bao2024GHZ,IBM2025GHZ}, the latter using $ZZ$ parity checks to detect preparation errors.
Measurement-based preparation with mid-circuit measurement and feedforward Pauli corrections~\cite{Baumer2024Dynamic} can instead produce GHZ resources in constant depth, eliminating the need for unitary CNOT-tree preparation circuits.
Our corridor preparation (Sec.~\ref{sec:arch_corridor}) is measurement-based and, before verification, constant-depth, adapted to the planar superconducting setting and to repeated extraction; its Pauli corrections are absorbed into the detector frame rather than applied by feedforward.

\section{Architecture}
\label{sec:arch}

\subsection{Circuit layout and qubit count}
\label{sec:arch_layout}

The full syndrome extraction circuit is parameterized by the inner-code distance $t$, the number of GHZ-state verification rounds $\rv$, the number of full outer-code syndrome measurement rounds $\ro$, and the circuit basis ($X$ or $Z$).  
Figure~\ref{fig:layout_d3} shows the stabilizer structure of our construction at $t = 3$: four rotated surface-code patches occupy the four quadrants of a square arrangement, with $N = 93$ physical qubits, $k = 2$ logical qubits, and code distance $d = 2t = 6$.
Two overlapping narrow ancilla \emph{corridors} of length $4t$ thread between the four patches. They are used to sequentially measure the outer-code stabilizer generators $G_X=\overline{XXXX}$ and $G_Z=\overline{ZZZZ}$.
Each corridor carries a $4t$-qubit GHZ resource state, and the corridor qubits next to the patches also serve as boundary measure qubits of the patches (Sec.~\ref{sec:arch_schedule}).

\begin{figure*}[t]
  \centering
  \includegraphics[width=0.9\textwidth]{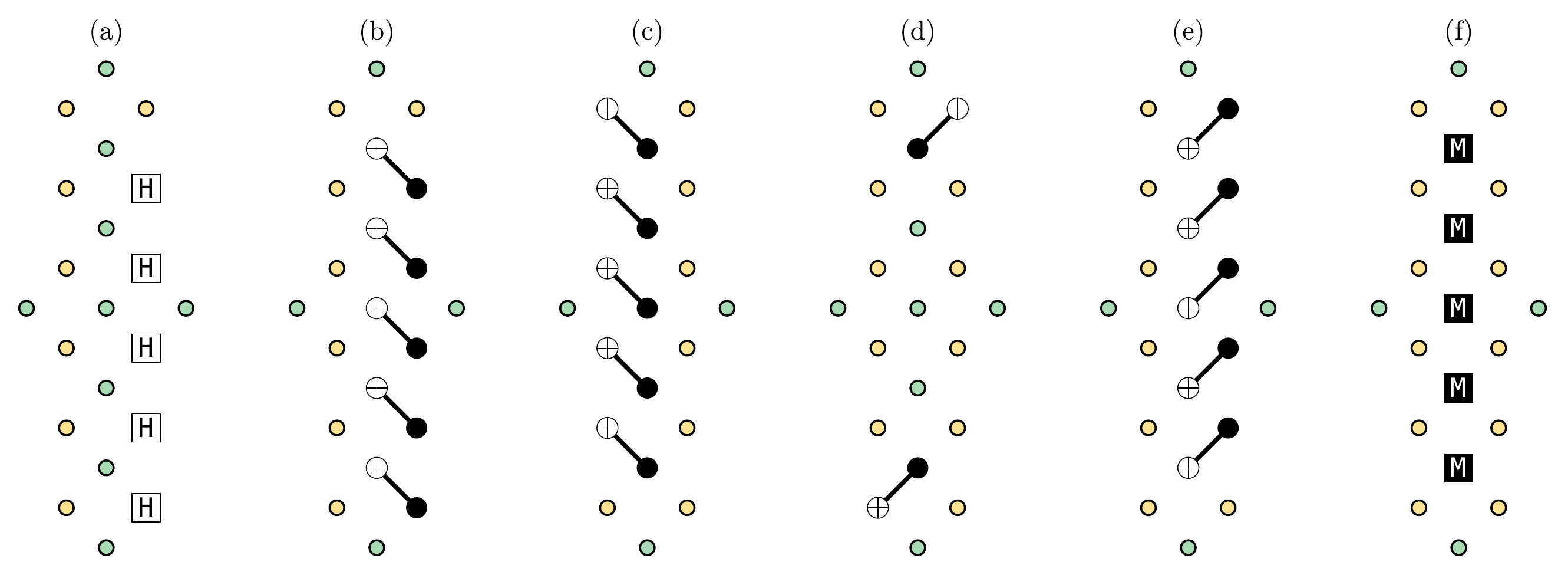}
  \caption{A single GHZ preparation round for patch distance $t = 3$, in six time steps (a)--(f). Neighboring corridor qubits are clustered into small GHZ states, intermediate qubits are measured, and the outcomes are absorbed virtually into the detector frame to align the GHZ branches.
  }
  \label{fig:prep}
\end{figure*}

\begin{figure*}[t]
  \centering
  \includegraphics[width=0.9\textwidth]{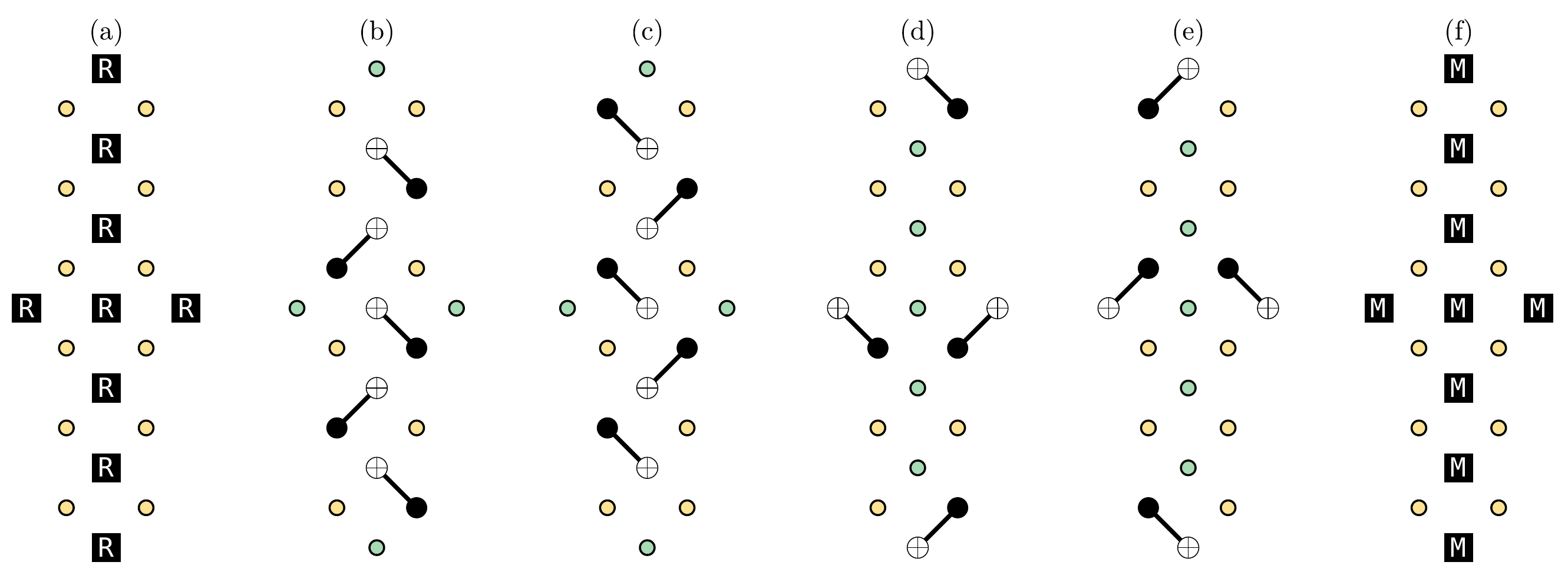}
  \caption{A single GHZ verification round for patch distance $t = 3$, in six time steps (a)--(f). Pairwise $ZZ$ parities along the corridor are measured by short-depth CNOT circuits using the ancillae on the corridor's center line, detecting $X$ errors on the GHZ resource state before the transversal CNOT gates couple it to the surface-code patches.
  }
  \label{fig:verf}
\end{figure*}

In a $Z$-basis circuit, the initial reset is followed by a single round of inner-codes' syndrome measurements and a $G_X$ measurement.  This preparation round is followed by $\ro$ outer-code rounds, each including one $G_Z$ and one $G_X$ measurement, followed by a final measurement of all data qubits in the $Z$ basis.  Each outer-code generator measurement consists of a single preparation round and $\rv$ verification rounds, each of the same depth and done concurrently with a round of the inner-code syndrome measurement, see Figs.~\ref{fig:prep} and \ref{fig:verf}.  Thus, preparation round ends at $\tau_0=\rv+2$, and the subsequent outer-code generators are measured at $\tau_1=2\rv+3$, $\tau_{2}=3\rv+4$, etc., with intervals of $\rv+1$.  The first $Z$-detector is placed at $\tau_1$, and the first $X$-detector at $\tau_2$; subsequent detectors are placed after each outer-code cycle of duration $2\rv+2$.

The $X$-basis circuits are constructed similarly, up to the basis for initial preparation / final measurements, and the order of outer-code generator measurements.  Namely, after the initial round of surface code measurements, inner-code measurements are combined with $G_Z$, $G_X$, $G_Z$, etc., done in this order, with the detectors placed at $\tau_1$, $\tau_2$, etc.  While the two circuits are not identical, they have the same repeat blocks in the bulk.  This choice is made to reduce the initial state preparation and measurement (SPAM) errors.  Indeed, if we were to measure $G_X$ first in the $X$-basis circuit, the first complete set of outer-code detectors would result only at $\tau_2$.

Combining four surface code patches with $N_{\rm surf}=2t^2-1$ physical qubits each with the $8t+1$ shared ancillary qubits required to measure the outer stabilizer generators (Sec.~\ref{sec:arch_schedule}) gives the physical footprint
\begin{equation}
N = 4(2t^2 - 1) + 8t + 1 = 2d^2 + 4d - 3,\qquad d = 2t.
\label{eq:total-qubits}
\end{equation}
In comparison, two independent distance-$d$ rotated surface-code patches use $2(2d^2-1)$ qubits, so at equal distance the stitched layout saves $2d^2-4d+1$ qubits.
The asymptotic encoding density,
\begin{equation}
\frac{k\, d^2}{N} \;=\; \frac{2(2t)^2}{8t^2 + 8t - 3} \;\xrightarrow{t\to\infty}\; 1,
\label{eq:efficiency}
\end{equation}
is twice the single-patch ratio $\tfrac{1}{2}$.

\subsection{GHZ corridor preparation and verification}
\label{sec:arch_corridor}

A single GHZ preparation round (Fig.~\ref{fig:prep}) constructs the resource state $\ket{\mathrm{GHZ}_{4t}}$ on the $4t$ corridor qubits using mid-circuit measurements~\cite{Baumer2024Dynamic}. The protocol prepares clusters of corridor qubits in smaller GHZ states, measures intermediate qubits in the appropriate fusion bases to form a GHZ line, and absorbs the resulting Pauli corrections virtually as detector-frame updates. The preparation circuit has constant depth, independent of $t$.

Preparation alone is not fault-tolerant: a single fault during preparation can leave a multi-qubit $X$ error on the corridor, which would propagate as a hook error of weight greater than one through the subsequent transversal CNOTs into data.
We follow each preparation by $\rv$ rounds of verification.
A verification round (Fig.~\ref{fig:verf}) measures pairwise $ZZ$ parities along the corridor by short-depth CNOTs into the ancillae on the corridor's center line; $X$ errors that arose during preparation or earlier verification rounds flip these parities, while $Z$ errors commute through the subsequent transversal CNOTs and contribute only to the readout. On the $\overline{Z}$-corridor the roles of $X$ and $Z$ are exchanged: the verification rounds measure pairwise $XX$ parities, which detect the $Z$ errors that would propagate into the data.

After $\rv$ verification rounds, the GHZ ancilla interacts transversally with the supporting data qubits of the four patches: CNOTs from corridor to data for the $\overline{X}$-corridor extraction, data to corridor for the $\overline{Z}$-corridor extraction. The corridor is then measured in the appropriate Pauli basis to read out the joint outer-stabilizer eigenvalue.

The verification depth sets a lower bound on the circuit distance.  For the circuits constructed with patch distances $t \le 10$ and $\rv \geq 2$ verification rounds, the distances computed by Stim~\cite{Gidney2021Stim} satisfy the empirical expression
\begin{equation}
d_{\rm circ} = \min(2t,\; \rv + 3).
\label{eq:dcirc}
\end{equation}
While the algorithm in Stim only guarantees the circuit distance for graph-like DEMs, we also use {\tt dist\_m4ri}~\cite{distm4ri} to calculate the distances.  For $d_{\rm circ}\le7$ Stim results are confirmed exactly with exhaustive cluster enumeration, while upper bounds for larger distances also agree with Eq.~(\ref{eq:dcirc}), see Appendix~\ref{app:circuit_distance} and Table~\ref{tab:circuit_distance}.  We conclude that the design circuit distance $2t$ is reached at $\rv = 2t-3$.
For smaller $\rv$, the shortest logical error is a time-like fault chain on the corridor: it repeats through the $\rv$ verification rounds at a single corridor site and is closed by three two-qubit-gate faults near the corridor intersection.
Being confined to a few corridor locations, this failure mode has a much smaller combinatorial weight than the spatial logical strings of the patches.
The optimal choice of $\rv$ is therefore set by the operating physical error rate: it suffices that the corridor contribution be well below the logical error rate of the bulk, as studied numerically in Sec.~\ref{sec:results_rv} (Fig.~\ref{fig:d5_rv_SI1000}).

\subsection{Reuse of boundary ancillae and scheduling}
\label{sec:arch_schedule}

The corridor qubits adjacent to each patch's boundary serve a dual role.
During the surface-code stabilizer cycle they participate as weight-two boundary syndrome ancillae for the inner patch; during the corridor cycle they participate as edge qubits of the GHZ resource state.
We schedule the two roles so that boundary stabilizer generators are temporarily disabled during the rounds in which the corresponding ancilla qubits are committed to the corridor, and re-enabled afterward.
This dual-use scheduling avoids a separate set of boundary ancillae along the corridors and keeps the total ancilla overhead at $8t+1$.

Overall, each outer-generator measurement occupies $\rv + 1$ inner cycles, which includes GHZ-state preparation, the $\rv$ verification rounds, transversal extraction, and the corridor readout.
A full outer-code measurement cycle---one round each of $G_X$ and $G_Z$ measurements---occupies $2(\rv + 1)$ inner-code cycles.

\section{Decoding}
\label{sec:decoding}

\label{sec:decoding_matching}
The decoder operates on the circuit-level detector error model (DEM) extracted by Stim~\cite{Gidney2021Stim} and uses minimum-weight perfect matching (MWPM)~\cite{Dennis2002,Higgott2025SparseBlossom}, which requires every fault mechanism to flip at most two detectors or to decompose into components that do.

The GHZ corridors add fault mechanisms that flip more than two detectors.
A fault on a corridor qubit during transversal extraction can spread onto the data qubits of adjacent patches; it then flips detectors in the corridor and in several patches, and, when the spread reaches a logical representative, a logical observable.
The corridor checks and verifier detectors are declared such that Stim decomposes every such mechanism into components that each flip at most two detectors.
The GHZ preparation outcomes define known Pauli frames on the corridor (Sec.~\ref{sec:arch_corridor}); rather than being applied as measurement-conditioned corrections, these frame bits are included directly in the detector declarations.
The resulting DEM is decoded by a single PyMatching call over all four patches~\cite{Higgott2022PyMatching,Higgott2025SparseBlossom}.
The outer-code detectors are nodes of the same matching graph, so error chains that cross patch boundaries are matched through them, and no separate outer-code decoding step is needed.
In the early stages of this work, before the circuits were made matchable, we validated them with the search-based decoder Tesseract~\cite{Tesseract2025}, which is slow for large distances.

Decomposition assigns independent weights to the components of a mechanism and therefore ignores their correlation.
Correlated matching in PyMatching (available since v2.3~\cite{Higgott2022PyMatching}) uses these correlations to reweight the matching graph after a first pass and then matches again~\cite{Fowler2013Correlated}.
We report both plain MWPM and correlated matching (Corr-MWPM) for the hardware data in Sec.~\ref{sec:experimental}.

\section{Numerical Results}
\label{sec:results}

\subsection{Simulation setup}
\label{sec:results_setup}

We simulate a memory experiment: the encoded state is initialized in a logical eigenstate, $\tau$ surface-code stabilizer rounds are executed, and at the end the data qubits are measured destructively in the corresponding basis.  For stitched code, $\tau$ corresponds to one full round of the outer code, $\tau=3\rv+4$, and we choose $\tau=3d$ for the surface code. The MWPM decoder of Sec.~\ref{sec:decoding} predicts the logical observables from the syndrome.  Since our $X$- and $Z$-basis circuits have very similar structure, with uniform depolarizing noise the corresponding error rates are nearly identical (data not shown).  Thus, all simulations are done for $Z$-basis circuits.

With two logical observables in the stitched code, a decoder reports the error probabilities $p_{\rm F,0}(\tau)$ and $p_{\rm F,1}(\tau)$ of each logical channel, and the block error probability $p_{\rm F,any}(\tau)$ corresponding to an error in either channel.  Empirically, the errors in the two logical channels have significant correlations.  To account for such correlations, we model the logical errors by a continuous-time stationary Markov chain model with states $\{00,01,10,11\}$, where a non-zero bit corresponds to an error in the corresponding channel, with bit-flip error rates $\mu_{01}$, $\mu_{10}$, and $\mu_{11}$.  The time is measured in the units of surface-code cycle duration, thus, e.g., the combined error rate $\mu_{\rm tot}=\mu_{01}+\mu_{10}+\mu_{11}$ is approximately equal to the block error probability per round, $\varepsilon_L$, see Appendix \ref{app:markov-chain} for the details.  With numerical data, we ignore the SPAM errors by setting initial polarizations in Eqs.~\eqref{eq:markov-p-0}, \eqref{eq:markov-p-1}, and \eqref{eq:markov-p-any} to $1$, and use the three measured values $p_{\rm F,0}(\tau)$, $p_{\rm F,1}(\tau)$, and $p_{\rm F,any}(\tau)$ to compute the three error rates.  This overestimates the error rates slightly but is sufficient to compare the error rates with those of the two logical qubits from two independent surface code patches.  

The full circuit is simulated in Stim~\cite{Gidney2021Stim} under two noise models: the SI1000 superconducting-inspired model of Gidney \emph{et al.}~\cite{GidneyNewmanFowlerBroughton2021} (two-qubit depolarizing channel with probability $p_2=p$ after a two-qubit gate, single-qubit depolarizing channel with probability $p_1=p/10$ after a single-qubit gate and on idle qubits, Pauli-$X$ channel with probability $p_{\rm meas}=5p$ before any measurement, and with probability $p_{\rm reset}=2p$ after any reset, and additional depolarization with probability $p_1'=2p$ on qubits idling during measurement or reset), and a uniform circuit-level depolarizing model with $p_2=p_1=p_{\rm meas}=p_{\rm reset}=p$ and $p_1'=0$.  Monte Carlo sampling uses the sinter harness with up to $4\times10^{9}$ shots per data point, and each point in Figs.~\ref{fig:SI1000_p_vs_LER}, \ref{fig:footprint_SI1000}, \ref{fig:STD_p_vs_LER}, and~\ref{fig:footprint_std} has at least $10^3$ logical errors, except four low-$p$ points limited by the $4\times10^{9}$-shot cap; uncertainties are 95\% Wilson score intervals, shown as shaded bands.
All shots are decoded; no postselection is applied.  We track both logical observables $L_0$ and $L_1$; the figures show the block error rate $\mu_{\rm tot}$.

\subsection{Effect of verification depth}
\label{sec:results_rv}

The verification depth $\rv$ controls a competition between hook suppression and costs of a longer outer-code round.
Increasing $\rv$ suppresses $X$ hook errors on the GHZ ancilla, raising the circuit distance toward its design value $2t$ [see Eq.~\eqref{eq:dcirc} and Table~\ref{tab:circuit_distance}].
Decreasing $\rv$ shortens the outer-code round, limiting (i) the number of inner-cycle $Z$ errors accumulated on the active GHZ ancilla, which add measurement noise to the readout; (ii)~the duration over which the boundary weight-2 surface-code detectors are disabled; and (iii) the inner-cycle latency between consecutive measurements of the other outer stabilizer.
The optimal $\rv$ therefore depends on $t$ and on the physical error rate $p$.

\begin{figure}[t]
  \centering
  \includegraphics[width=\linewidth]{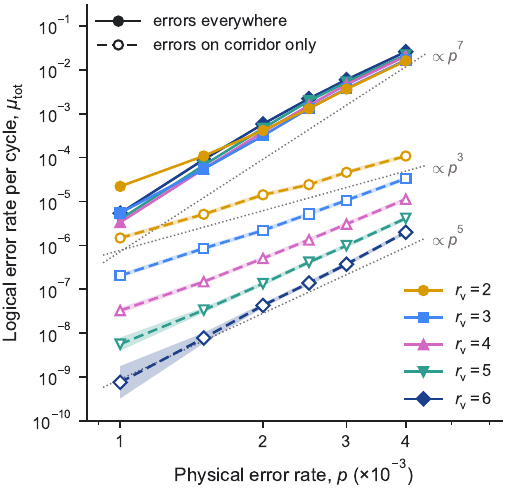}
  \caption{Logical error rate $\mu_{\rm tot}$ per surface-code cycle (Sec.~\ref{sec:results_setup}) versus physical error rate $p$ for patch distance $t=7$ under SI1000 noise, with $\rv=2$--$6$ GHZ verification rounds (Sec.~\ref{sec:arch_corridor}). Solid curves: errors present on all operations ($p_{\rm bulk}=p_{\rm GHZ}$). Dashed curves: errors present on the GHZ corridor only ($p_{\rm bulk}=0$). Dotted lines are guides to the eye $\propto p^3$, $p^5$, $p^7$. Bands are 95\% Wilson intervals.}
  \label{fig:d5_rv_SI1000}
\end{figure}

As $p$ decreases, the optimal $\rv$ increases (Fig.~\ref{fig:d5_rv_SI1000}): at $t=7$ it moves from $\rv=2$--$3$ at $p=4\times10^{-3}$ to $\rv=4$ at $p=10^{-3}$, where the larger circuit distance pays off.
At intermediate and large $p$, smaller $\rv$ wins because the prefactor is reduced.
For every data point in our subsequent scaling figures we choose $\rv$ to minimize the joint logical error rate, requiring that the choice be simultaneously optimal (or near-optimal) for both logical qubits.

\subsection{Distance scaling}
\label{sec:results_scaling}

\begin{figure}[t]
  \centering
  \includegraphics[width=\linewidth]{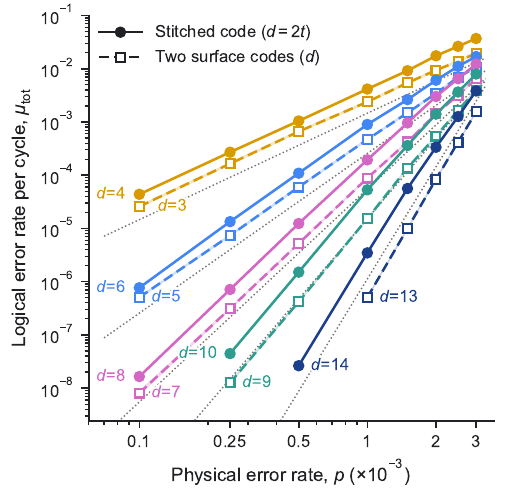}
  \caption{Logical error rate $\mu_{\rm tot}$ per surface-code cycle versus physical error rate $p$ under the SI1000 noise model, for the stitched code of distance $d=2t$ (filled circles, solid lines) and two independent rotated surface-code patches of distance $d$ (open squares, dashed lines); numbers on the curves give $d$. Shown are $d=4,6,8,10,14$ for the stitched code and $d=3,5,7,9,13$ for the surface code. At each stitched point $\rv$ is the value with the lowest joint logical error rate, choosing among statistically tied values the one with the most similar $L_0$ and $L_1$ rates (Sec.~\ref{sec:results_rv}). Dotted lines are guides to the eye $\propto p^{t}$.}
  \label{fig:SI1000_p_vs_LER}
\end{figure}

Figure~\ref{fig:SI1000_p_vs_LER} compares the logical error rate of the stitched code against two independent rotated surface-code patches of varying distance under SI1000; Fig.~\ref{fig:STD_p_vs_LER} in Appendix~\ref{app:uniform} repeats the comparison under the uniform depolarizing model.
In both models the stitched code's logical error rate scales as $p^{t}$ (slope $t$ on log-log axes), consistent with code distance $d=2t$.

\subsection{Footprint comparison}
\label{sec:results_footprint}

At equal distance, the stitched layout uses fewer qubits (Sec.~\ref{sec:arch_layout}), but the more relevant figure of merit in practice is the number of physical qubits needed to reach a target logical error rate per round, $\varepsilon_L$, at fixed $p$.

\begin{figure}[t]
  \centering
  \includegraphics[width=\linewidth]{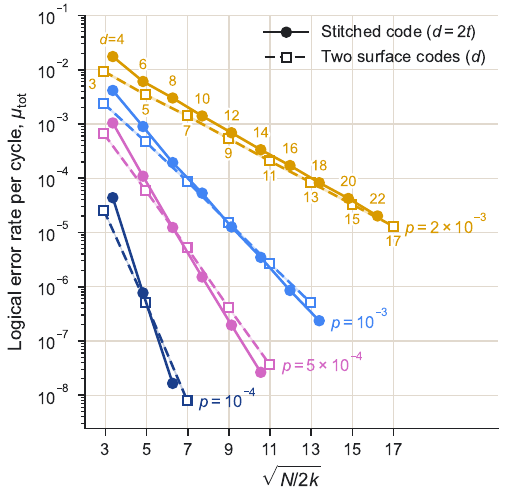}
  \caption{Logical error rate $\mu_{\rm tot}$ per surface-code cycle versus $\sqrt{N/2k}$ under the SI1000 noise model, where $N$ is the total number of qubits and $k=2$ the number of logical qubits, for the stitched code (filled circles, solid lines) and two surface-code patches (open squares, dashed lines) at the physical error rates $p$ written next to the curves; a surface code of distance $d$ sits at $\sqrt{N/2k}\approx d$. Data of Fig.~\ref{fig:SI1000_p_vs_LER}, all simulated distances.}
  \label{fig:footprint_SI1000}
\end{figure}

Under SI1000 noise (Fig.~\ref{fig:footprint_SI1000}), the stitched code reaches a given $\varepsilon_L$ with fewer qubits than two surface-code patches once the codes are sufficiently large: the crossover occurs at $N\approx 157$ for $p=10^{-4}$, $N\approx 237$ for $p=5\times10^{-4}$, and $N\approx 333$ for $p=10^{-3}$. At $p=(2\text{--}3)\times10^{-3}$, by contrast, its logical error rate at $N\approx 10^3$ remains 10--14\% higher than that of two surface-code patches.

The same comparison under the uniform depolarizing model is given in Appendix~\ref{app:uniform}.

The footprint advantage is therefore not universal but depends on $p$ and on the target error rate.
Asymptotically the stitched code uses half as many physical qubits per logical qubit as two independent surface patches of matched code distance [Eq.~\eqref{eq:efficiency}]; this advantage is realized in the footprint comparison once $t$ is large enough that the corridor-noise prefactor is amortized.
The crossover lies at physical error rates within reach of near-term superconducting hardware. The physical error rates of the experiment reported in Sec.~\ref{sec:experimental} (Fig.~\ref{fig:hardware_err}) lie above this crossover.

\section{Experimental Results}
\label{sec:experimental}
\begin{figure*}[t]
    \centering
    \includegraphics[width=0.65\textwidth]{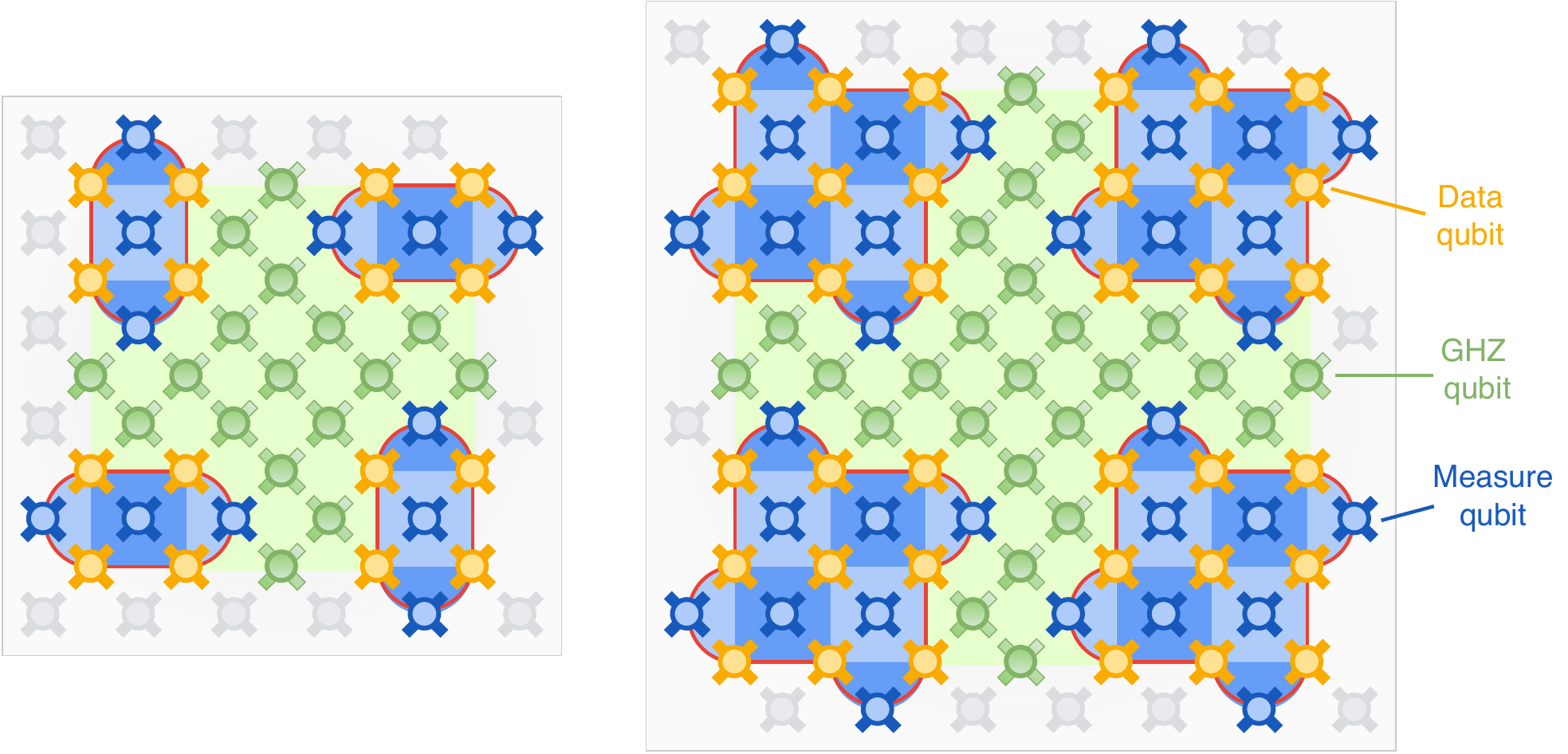}%
    \caption{Layout of the stitched-code experiments on the 105-qubit Willow processor, $d=4$ ($t=2$, left) and $d=6$ ($t=3$, right), drawn to the same scale. Each measure qubit (blue) is associated with an inner-patch stabilizer (blue tile). Red outlines indicate the four rotated surface-code patches, containing 16 and 36 data qubits in total (gold). GHZ overhead qubits (green) are located in the corridors between the patches and are used to measure the outer-code stabilizer generators across the four patches. Qubits colored gray are not used.}
    \label{fig:hardware_grid}
\end{figure*}

\begin{figure}[htbp]
    \centering
    \includegraphics[width=0.359\textwidth]{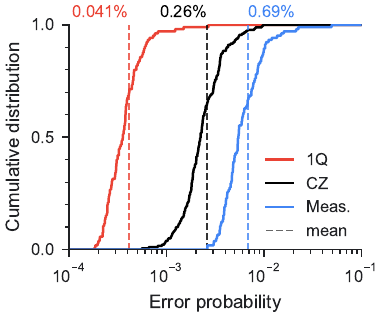}
    \caption{Cumulative distributions of physical error probabilities measured on the processor during the experiment: red denotes single-qubit gate Pauli errors, black CZ Pauli errors, and blue measurement-identification error; dashed vertical lines mark the means, whose values are given above the panel.}
    \label{fig:hardware_err}
\end{figure}

In the experiment we executed the full stitched-code circuits of two sizes, including repeated GHZ-corridor extraction, on a superconducting quantum processor, and verified that the logical error rate decreases as the code distance increases.
We used a 105-qubit Willow superconducting processor, the device from the family used for the below-threshold surface-code memory experiments of Ref.~\cite{Google2025BelowThreshold}, using reinforcement-learning-based calibration methods similar to those used in Ref.~\cite{sivak2025reinforcement}.
The hardware circuits apply data-qubit leakage removal (DQLR)~\cite{Miao2023Leakage} at the boundaries of each GHZ block.
The device embedding used for the stitched layout and the calibrated physical error distributions are shown in Figs.~\ref{fig:hardware_grid} and~\ref{fig:hardware_err}.

The stitched-code layout differs in one important respect from the native layout used in standard rotated surface-code memory experiments~\cite{Fowler2012PRA,Google2025BelowThreshold}. In the native layout, data and syndrome qubits form a bipartite pattern: the syndrome-qubit sublattice is optimized for repeated measurement and reset, while the data-qubit sublattice is optimized for coherent storage followed by a single final readout~\cite{Google2025BelowThreshold}. In the stitched circuits, by contrast, some qubits that would ordinarily serve as data qubits are repurposed as corridor ancillae and must therefore be measured and reset in every outer-code extraction round. This makes the hardware implementation more demanding than a native surface-code memory experiment: although the circuit uses the same processor primitives, it requires repeated measurement and reset on qubits that are not optimized for these operations.

\subsection{Experimental setup}
\label{sec:experimental_setup}

We ran two code sizes, with the verification depth chosen for the lowest logical error rate at the hardware operating point (Sec.~\ref{sec:results_rv}).
The first code with $d=4$ uses patch distance $t=2$ and $\rv=1$ verification round, requiring four inner surface-code cycles per full outer-code measurement cycle.
The second code with $d=6$ uses patch distance $t=3$ and $\rv=2$ verification rounds, requiring six inner cycles per full outer-code cycle.  Notice that the circuit distance for the latter code is $d_{\rm circ}=5$, smaller than the code distance $d=6$.

For each code, we ran memory experiments in two bases: in the $ZZ$-basis experiment both logical qubits are prepared and measured in the $Z$ basis, and in the $XX$-basis experiment both are prepared and measured in the $X$ basis.
For each instance and basis, we executed circuits with $\ro\in\{1,\dots,8\}$ full outer-code rounds.
These correspond to $\tau\in\{7,11,\dots,35\}$ inner surface-code cycles for $t=2$ and $\tau\in\{10,16,\dots,52\}$ for $t=3$.

As discussed in Sec.~\ref{sec:decoding_matching}, we use the PyMatching decoder with the detector error model constructed for each circuit with SI1000 error model priors at $p=10^{-3}$ (Sec.~\ref{sec:results_setup}).
We used both regular MWPM and correlated matching (Corr-MWPM), applied to the same sets of measured data.  For each data set, the latter decoder shows better accuracy.  Thus, all the plots show only the Corr-MWPM data.

\begin{figure*}[htbp]
    \includegraphics[width=\textwidth]{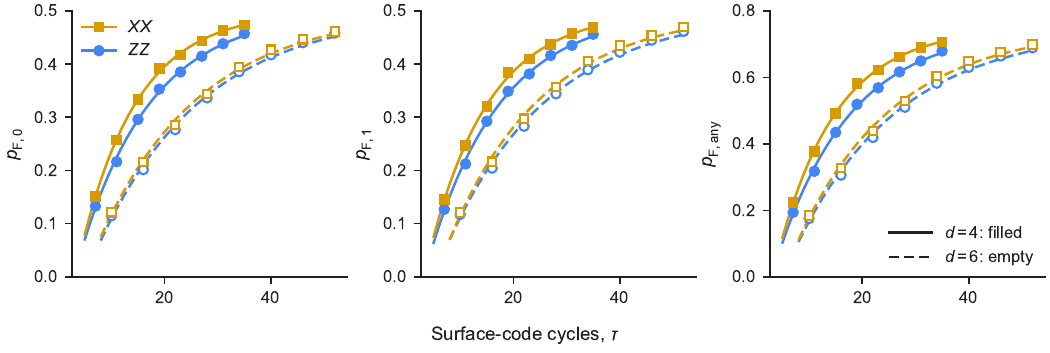}
    \caption{Experimental logical error probabilities (using correlated matching for decoding) as a function of the number of surface-code cycles $\tau$ in the circuit: $p_{\rm F,0}$ (left), $p_{\rm F,1}$ (middle), and the block error probability $p_{\rm F,any}$ (right). Squares (circles) show results for $XX$ ($ZZ$) memory experiments; filled markers and solid lines for the $d=4$ code, empty markers and dashed lines for the $d=6$ code. Lines show the corresponding values from the Markov chain model using three initial values and three error rates as fitting parameters, see App.~\ref{app:markov-chain}.  Data from each experiment were fitted separately, see Tab.~\ref{tab:hardware_ler_summary}. Error bars are 95\% Wilson intervals of the shot counts and are smaller than the symbols.}
    \label{fig:experimental_observables}
\end{figure*}

For each round, the decoder returns two bits corresponding to the error status in each observable.  Thus, for each set of measurements, four probabilities are calculated, consistent with the number of the outcomes.  We report the marginal error probabilities for the logical variables, $p_{\rm F,0}=P_{01}+P_{11}$ and $p_{\rm F,1}=P_{10}+P_{11}$, as well as the block error probability $p_{\rm F,any}=P_{01}+P_{10}+P_{11}$.  These results are shown in Fig.~\ref{fig:experimental_observables} with symbols, separately for each distance and circuit basis.  The left and center panels are for the two logical observables, $p_{\rm F,0}$ and $p_{\rm F,1}$, while the right panel shows the block error rate $p_{\rm F,any}$.
Each circuit was sampled with $200{,}000$ shots.  The statistical errors are smaller than the symbol size.

\subsection{Modeling of error rates}
\label{sec:extended_cycles}

Unlike the reference model of two surface code patches, the errors for the two logical qubits in the stitched code are not independent.   While a single error rate can be used to fit the data for either of the logical observables separately (as is commonly done for the surface code), this fails for the block error rate $p_{\rm F,any}$.  To analyze the data in Fig.~\ref{fig:experimental_observables}, we used a continuous-time four-state Markov chain model with three distinct error rates: independent flip rates $\mu_{01}$, $\mu_{10}$ of each logical observable, and $\mu_{11}$, the rate of correlated flips of both, see Appendix \ref{app:markov-chain}.  Since the initial preparation time and the outer-code cycle both take several inner-code cycles (see Sec.~\ref{sec:arch_layout}), such a model is not expected to be accurate at a scale of a single cycle.  In particular, the initial polarizations at $\tau=0$ in Eqs.~\eqref{eq:markov-p-0}, \eqref{eq:markov-p-1}, \eqref{eq:markov-p-any} are considered as free parameters, and are even allowed to take nonphysical values in such a fit.  Indeed, while a continuous-time Markov chain is equivalent to a discrete model for a periodic process like a measurement cycle, the state preparation part of the circuit warrants a transition matrix different from that in the rest of the circuit.  Further, the initial polarizations $S_0(0)$, $S_1(0)$, and $S_{\rm parity}(0)$ include the SPAM errors which are not guaranteed to be identical.

Thus, for each set of average probabilities $P_{\rm F,0}(\tau)$, $P_{\rm F,1}(\tau)$, and $P_{\rm F,any}(\tau)$, we used Eqs.~\eqref{eq:markov-p-0}, \eqref{eq:markov-p-1}, \eqref{eq:markov-p-any} to fit the observable flip probabilities in Fig.~\ref{fig:experimental_observables} (and similar data from regular MWPM decoder), using the three flip rates and three initial polarizations as independent free parameters.  The data from each experiment computed with each decoder were used for separate fits.  The corresponding process rates are shown in Table \ref{tab:markov_generator_rates}.  The fitted curves are shown in Fig.~\ref{fig:experimental_observables} as solid and dashed lines for $d=4$ and $d=6$, respectively.  

\begin{table}[htbp]
\centering
\footnotesize
\begin{ruledtabular}
\begin{tabular}{lccccc}
\textbf{Decoder} & \textbf{Dist} & \textbf{Basis} & \textbf{\boldmath$\mu_{01}$} & \textbf{\boldmath$\mu_{10}$} & \textbf{\boldmath$\mu_{11}$} \\
\midrule
MWPM & $d=4$ & $ZZ$ & $2.065(14)$ & $2.032(14)$ & $2.123(14)$ \\
MWPM & $d=4$ & $XX$ & $2.933(23)$ & $2.605(23)$ & $2.610(23)$ \\
MWPM & $d=6$ & $ZZ$ & $1.328(9)$ & $1.420(9)$ & $1.354(9)$ \\
MWPM & $d=6$ & $XX$ & $1.432(10)$ & $1.636(10)$ & $1.495(10)$ \\
\midrule
Corr-MWPM & $d=4$ & $ZZ$ & $1.686(11)$ & $1.667(11)$ & $2.029(12)$ \\
Corr-MWPM & $d=4$ & $XX$ & $2.384(16)$ & $2.132(16)$ & $2.261(16)$ \\
Corr-MWPM & $d=6$ & $ZZ$ & $1.221(7)$ & $1.305(8)$ & $1.275(8)$ \\
Corr-MWPM & $d=6$ & $XX$ & $1.285(8)$ & $1.496(9)$ & $1.323(8)$ \\
\end{tabular}
\end{ruledtabular}
\caption{Rates (percent per surface code cycle) of the four-state Markov model used to fit the data in Fig.~\ref{fig:experimental_observables}, see Appendix \ref{app:markov-chain}.  The parameters $\mu_{01}$, $\mu_{10}$ are independent flip rates of each logical observable, while $\mu_{11}$ is the rate of correlated flips of both. The data for different distances and different circuit bases are fitted separately. Parenthetical uncertainties are one-standard-deviation uncertainties in the last quoted digits.}
\label{tab:markov_generator_rates}
\end{table}

\begin{table}[htbp]
\centering
\footnotesize
\begin{ruledtabular}
\begin{tabular}{@{}l c c c c@{}}
\textbf{Decoder} & \textbf{Error} & \textbf{$d=4$} & \textbf{$d=6$} & \textbf{$\Lambda_{4\to6}$} \\
\midrule
\multirow{3}{*}{MWPM}
& \(L_0\) & \(4.632(15)\%\) & \(2.727(8)\%\) & \(1.699(7)\) \\
& \(L_1\) & \(4.469(14)\%\) & \(2.866(8)\%\) & \(1.559(7)\) \\
& Any & \(6.845(15)\%\) & \(4.209(8)\%\) & \(1.626(5)\) \\
\midrule
\multirow{3}{*}{Corr-MWPM}
& \(L_0\) & \(4.008(12)\%\) & \(2.488(6)\%\) & \(1.611(6)\) \\
& \(L_1\) & \(3.884(11)\%\) & \(2.628(7)\%\) & \(1.478(6)\) \\
& Any & \(5.836(12)\%\) & \(3.850(7)\%\) & \(1.516(4)\) \\
\end{tabular}
\end{ruledtabular}
\caption{Hardware logical error probabilities per surface-code cycle. The columns use the stitched-code distance $d=2t$, with patch distances $t=2$ and $t=3$ corresponding to $d=4$ and $d=6$, respectively. All entries use the Markov-chain fit described in Sec.~\ref{sec:extended_cycles}, averaged over the $ZZ$ and $XX$ experiments; parenthetical uncertainties are one-standard-deviation sampling uncertainties in the last quoted digits, from the likelihood curvature of the fit. Rows labeled $L_0$ and $L_1$ give the per-cycle error probabilities $\varepsilon_{0}$ and $\varepsilon_{1}$ of the two logical observables, and ``Any'' the per-cycle block error probability $\varepsilon_{L}$, computed from the fitted rates of Table~\ref{tab:markov_generator_rates} (Appendix~\ref{app:markov-chain}).}
\label{tab:hardware_ler_summary}
\end{table}

\begin{figure*}[tbp]
    \centering
    \includegraphics[width=\textwidth]{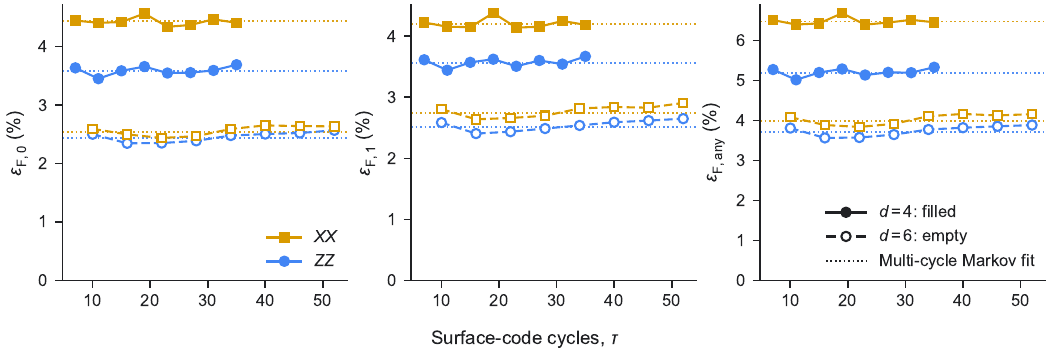}
    \caption{Single-point Markov-chain estimates of the per-cycle logical error probabilities $\varepsilon_{0}$, $\varepsilon_{1}$, and $\varepsilon_L$ (Appendix~\ref{app:markov-chain}) for the Corr-MWPM data at every measured depth $\tau$, with the fitted initial polarizations of Fig.~\ref{fig:experimental_observables} used as the reference at $\tau=0$. Squares show the data for $XX$ and circles $ZZ$ memory experiments; filled markers and solid lines correspond to $d=4$, empty markers and dashed lines to $d=6$.  The dotted lines show the  corresponding values computed from six-parameter Markov-chain fits of each experiment (also listed in Table~\ref{tab:markov_generator_rates}). Error bars at 95\% Wilson intervals are smaller than the symbol size.}
    \label{fig:corr_markov_moving_average}
\end{figure*}

As an additional finite-duration diagnostic, we also computed single-point rate estimates from the same data.  Namely, for each $\tau$, we used Eqs.~\eqref{eq:markov-p-0}, \eqref{eq:markov-p-1}, \eqref{eq:markov-p-any} (with the initial polarization values from the overall fit) to compute the rates $\mu_0$, $\mu_1$, and $\mu_2$ from the three values $p_{\rm F,0}(\tau)$, $p_{\rm F,1}(\tau)$, and $p_{\rm F,any}(\tau)$.  The results are shown in Fig.~\ref{fig:corr_markov_moving_average}.  The single-point rates thus computed do not show a significant change with $\tau$, i.e., the observed logical error rates remain more or less constant up to the maximum duration of the experiment, $\tau_{\rm max}=52$ surface code cycles.

\subsection{Logical error probabilities per cycle}
\label{sec:experimental_error_analysis}
For quantum memory based on surface codes, a de facto standard measure used to compare 
experiments of different duration is the average logical error probability per cycle, $\varepsilon$~\cite{Google2021Exponential,Google2025BelowThreshold}.

With two logical qubits, there are three related probabilities (per surface-code cycle) that can be computed for each experiment: the marginal error probability in each logical channel, $\varepsilon_0$ and $\varepsilon_1$, and the block error probability, $\varepsilon_L$.  Since we use the duration of the surface code cycle as the time unit, to linear order in small error rates, these probabilities are numerically close to the respective marginal error rates $\mu_0=\mu_{01}+\mu_{11}$, $\mu_1=\mu_{10}+\mu_{11}$, and the total error rate $\mu_{\rm tot}=\mu_{01}+\mu_{10}+\mu_{11}$.  The full expressions for the three probabilities are given in Eq.~\eqref{eq:epsilon-01L}.  

We used these equations to compute per-cycle logical error probabilities.  The resulting basis-averaged quantities are shown in Table~\ref{tab:hardware_ler_summary}, along with the corresponding improvement factors $\Lambda_{4\to6}\equiv \varepsilon^{(d=4)}/\varepsilon^{(d=6)}$.
In particular, for plain MWPM, increasing the code distance from $d=4$ to $d=6$ reduces the block error probability per cycle $\varepsilon_{L}$ from $6.845(15)\%$ to $4.209(8)\%$, corresponding to an improvement factor $\Lambda_{4\to6}=1.626(5)$.
Correlated matching lowers the error probability by $15\%$ at $d=4$, to $5.836(12)\%$, and by $9\%$ at $d=6$, to $3.850(7)\%$, which results in a smaller $\Lambda_{4\to6}=1.516(4)$.

\subsection{Differences with surface code experiments}

One immediate observation from the experimental plots in Figs.~\ref{fig:experimental_observables} and \ref{fig:corr_markov_moving_average} is that the $X$-basis circuits have higher logical error rates than the $Z$-basis circuits, with the difference most pronounced at $d=4$.  This is confirmed by the per-basis fitted rates listed in Table~\ref{tab:markov_generator_rates}.  Given that the $X$- and $Z$-basis circuits are nearly identical, we explain this asymmetry by low-frequency phase noise which cannot be entirely eliminated by the device calibration.  We mention in passing that phase noise was also present in the surface-code memory experiment done on a similar Willow device~\cite{Google2025BelowThreshold}.  The basis asymmetry was suppressed by using an $XZZX$ version of the surface codes~\cite{BonillaAtaides2021XZZX}, unlike the CSS codes we use in this work.

Another important observation is that the outer-code detectors, even though they are not used frequently, are absolutely essential to achieve the small error rates for stitched codes.
These detectors show much higher firing rates than the regular surface-code detectors [see Fig.~\ref{fig:det_frac} and Appendix~\ref{app:detector_diagnostics}], as would be expected from the large physical weight $4t$ of the outer-code stabilizer generators.
Indeed, in the $d=6$ hardware experiment they fire with probability close to $1/2$, yet flipping the value of any one of the primary-basis outer-code detectors 
changes a decoded logical in about $80\%$ of the shots, and in these shots it is the measured value, rather than its flip, that gives the correct logical outcome $95\%$ of the time (as computed for the experiment with $\ro=1$, where each outer-code detector is present only once, see Appendix~\ref{app:outer_role}).

The high firing rate of the outer-code detectors also delimits the role of the corridor GHZ gadgets.
A memory experiment can afford such a rate: the outer code only has to catch the rare logical failures of individual patches, and it has no deadline, so each further outer-code round makes such a failure easier to tell from a misread.
In contrast, a parity measurement between patches must settle the eigenvalue of an outer-code generator by a given time.  Here, one outer-code round per $2(\rv+1)$ surface-code cycles is way too slow for most applications.  Such a sparsity of undetectable time-like outer-code error chains is quantified in 
Appendix~\ref{app:stability} with a stability circuit simulation~\cite{Gidney2022Stability}: at $p=10^{-3}$, the stitched code with $t=3$ needs about $80$ cycles to determine $G_X$ with an error probability of $10^{-3}$, while a surface code of the same distance about $5$ cycles.
For extracting parities, GHZ corridors are thus far slower than lattice surgery; logical operations on stitched codes should leave the corridors aside and instead go through patch boundaries or use code automorphisms.

\section{Outlook and Conclusion}
\label{sec:outlook}

We have introduced a stitched concatenated memory that combines four rotated surface-code patches with the $[\,\![4,2,2]\,\!]$ outer detection code in a strictly planar nearest-neighbor layout. The construction encodes two logical qubits at code distance $d=2t$ while using a physical footprint $N=2d^2+4d-3$, and its verified GHZ corridors measure the two weight-$4t$ outer stabilizer generators without lattice surgery, long-range gates, or workspace patches. Circuit-level simulations show the expected distance-$2t$ scaling (Figs.~\ref{fig:SI1000_p_vs_LER} and~\ref{fig:STD_p_vs_LER}), and the hardware experiment on a 105-qubit Willow processor shows that the same architecture can be executed on a present superconducting device, with the logical error per cycle decreasing from $d=4$ to $d=6$ in both memory bases (Table~\ref{tab:hardware_ler_summary}).

The main practical tradeoff is the corridor prefactor: the asymptotic factor-of-two gain in qubits per logical qubit is realized below a crossover physical error rate that lies within reach of near-term superconducting hardware. The stitched layout also packs both logical qubits into a single, nearly square region, whereas two independent patches of comparable distance form a rectangle. In situations where availability, processor tiles, packaging, control wiring, or modular interconnects favor square regions, this geometry is an advantage independent of qubit count.

Several directions remain open. Hardware-specific optimization of the corridor readout/reset path, leakage-removal scheduling, and calibration of the dual-use boundary ancillae should reduce the corridor prefactor. More broadly, the GHZ-corridor gadget is not specific to the $[\,\![4,2,2]\,\!]$ code: it suggests a way to realize other small outer codes whose stabilizer generators can be routed through planar corridors. Identifying such codes, comparing their layout and decoder tradeoffs, and developing logical operations on and between stitched blocks given their slower accumulation of time-like distance, are natural next steps toward using compact concatenated modules as building blocks for larger fault-tolerant architectures.


\begin{acknowledgments}
We thank Adam Zalcman for reviewing the manuscript, Vlad Sivak for help setting up RL calibration, and Kaavya Sahay and Junichi Haruna for conversations on decoding. We also thank Oscar Higgott, Bryce Kobrin, Fengyi Li, Matt McEwen, Michael Newman, and Kevin Satzinger for useful conversations. We thank Agustin Di Paolo and Dvir Kafri for their mentorship and support of M.A.S. at Google Quantum AI. We thank Google Quantum AI for providing the research environment that made this work possible.
\end{acknowledgments}

\section*{Data availability}
The circuit generator, the simulation data under the SI1000 and uniform depolarizing noise models (stitched code and surface-code baseline, every simulated verification depth, and the verification-depth scan of the $t=7$ code), and the per-circuit logical-outcome counts of the hardware experiment are available at \url{https://github.com/QEC-pages/Stitched-Surface-Code}. The raw detection events and measured observables of all 32 hardware circuits, with the detector error models and decoder outputs used in the analysis, are provided as a release asset of the same repository.

\section*{Author contributions}
M.A.S. led the project in close collaboration with L.P.P., who supervised it throughout. C.G. contributed ideas and constructive criticism in early discussions, later suggested the footprint comparison, and made a strong case for redefining the detectors so that the circuits can be decoded by standard matching, which proved correct. A.B. helped set up the initial experiment. A.K. and A.M. contributed to the state-of-the-art calibration of the device. N.N. advised on fitting and data taking. L.M. helped adapt the hardware circuits to mitigate low-frequency noise.

\bibliography{references}

\begin{thebibliography}{56}%
\makeatletter
\providecommand \@ifxundefined [1]{%
 \@ifx{#1\undefined}
}%
\providecommand \@ifnum [1]{%
 \ifnum #1\expandafter \@firstoftwo
 \else \expandafter \@secondoftwo
 \fi
}%
\providecommand \@ifx [1]{%
 \ifx #1\expandafter \@firstoftwo
 \else \expandafter \@secondoftwo
 \fi
}%
\providecommand \natexlab [1]{#1}%
\providecommand \enquote  [1]{``#1''}%
\providecommand \bibnamefont  [1]{#1}%
\providecommand \bibfnamefont [1]{#1}%
\providecommand \citenamefont [1]{#1}%
\providecommand \href@noop [0]{\@secondoftwo}%
\providecommand \href [0]{\begingroup \@sanitize@url \@href}%
\providecommand \@href[1]{\@@startlink{#1}\@@href}%
\providecommand \@@href[1]{\endgroup#1\@@endlink}%
\providecommand \@sanitize@url [0]{\catcode `\\12\catcode `\$12\catcode
  `\&12\catcode `\#12\catcode `\^12\catcode `\_12\catcode `\%12\relax}%
\providecommand \@@startlink[1]{}%
\providecommand \@@endlink[0]{}%
\providecommand \url  [0]{\begingroup\@sanitize@url \@url }%
\providecommand \@url [1]{\endgroup\@href {#1}{\urlprefix }}%
\providecommand \urlprefix  [0]{URL }%
\providecommand \Eprint [0]{\href }%
\providecommand \doibase [0]{https://doi.org/}%
\providecommand \selectlanguage [0]{\@gobble}%
\providecommand \bibinfo  [0]{\@secondoftwo}%
\providecommand \bibfield  [0]{\@secondoftwo}%
\providecommand \translation [1]{[#1]}%
\providecommand \BibitemOpen [0]{}%
\providecommand \bibitemStop [0]{}%
\providecommand \bibitemNoStop [0]{.\EOS\space}%
\providecommand \EOS [0]{\spacefactor3000\relax}%
\providecommand \BibitemShut  [1]{\csname bibitem#1\endcsname}%
\let\auto@bib@innerbib\@empty
\bibitem [{\citenamefont {Fowler}\ \emph {et~al.}(2012)\citenamefont {Fowler},
  \citenamefont {Mariantoni}, \citenamefont {Martinis},\ and\ \citenamefont
  {Cleland}}]{Fowler2012PRA}%
  \BibitemOpen
  \bibfield  {author} {\bibinfo {author} {\bibfnamefont {A.~G.}\ \bibnamefont
  {Fowler}}, \bibinfo {author} {\bibfnamefont {M.}~\bibnamefont {Mariantoni}},
  \bibinfo {author} {\bibfnamefont {J.~M.}\ \bibnamefont {Martinis}},\ and\
  \bibinfo {author} {\bibfnamefont {A.~N.}\ \bibnamefont {Cleland}},\
  }\bibfield  {title} {\bibinfo {title} {Surface codes: Towards practical
  large-scale quantum computation},\ }\href
  {https://doi.org/10.1103/physreva.86.032324} {\bibfield  {journal} {\bibinfo
  {journal} {Phys. Rev. A}\ }\textbf {\bibinfo {volume} {86}},\ \bibinfo
  {pages} {032324} (\bibinfo {year} {2012})},\ \Eprint
  {https://arxiv.org/abs/1208.0928} {arXiv:1208.0928} \BibitemShut {NoStop}%
\bibitem [{\citenamefont {Dennis}\ \emph {et~al.}(2002)\citenamefont {Dennis},
  \citenamefont {Kitaev}, \citenamefont {Landahl},\ and\ \citenamefont
  {Preskill}}]{Dennis2002}%
  \BibitemOpen
  \bibfield  {author} {\bibinfo {author} {\bibfnamefont {E.}~\bibnamefont
  {Dennis}}, \bibinfo {author} {\bibfnamefont {A.}~\bibnamefont {Kitaev}},
  \bibinfo {author} {\bibfnamefont {A.}~\bibnamefont {Landahl}},\ and\ \bibinfo
  {author} {\bibfnamefont {J.}~\bibnamefont {Preskill}},\ }\bibfield  {title}
  {\bibinfo {title} {Topological quantum memory},\ }\href
  {https://doi.org/10.1063/1.1499754} {\bibfield  {journal} {\bibinfo
  {journal} {J. Math. Phys.}\ }\textbf {\bibinfo {volume} {43}},\ \bibinfo
  {pages} {4452} (\bibinfo {year} {2002})},\ \Eprint
  {https://arxiv.org/abs/quant-ph/0110143} {arXiv:quant-ph/0110143}
  \BibitemShut {NoStop}%
\bibitem [{\citenamefont {Kitaev}(2003)}]{Kitaev2003}%
  \BibitemOpen
  \bibfield  {author} {\bibinfo {author} {\bibfnamefont {A.~Y.}\ \bibnamefont
  {Kitaev}},\ }\bibfield  {title} {\bibinfo {title} {Fault-tolerant quantum
  computation by anyons},\ }\href
  {https://doi.org/10.1016/s0003-4916(02)00018-0} {\bibfield  {journal}
  {\bibinfo  {journal} {Ann. Phys.}\ }\textbf {\bibinfo {volume} {303}},\
  \bibinfo {pages} {2} (\bibinfo {year} {2003})},\ \Eprint
  {https://arxiv.org/abs/quant-ph/9707021} {arXiv:quant-ph/9707021}
  \BibitemShut {NoStop}%
\bibitem [{\citenamefont {Horsman}\ \emph {et~al.}(2012)\citenamefont
  {Horsman}, \citenamefont {Fowler}, \citenamefont {Devitt},\ and\
  \citenamefont {Van~Meter}}]{Horsman2012}%
  \BibitemOpen
  \bibfield  {author} {\bibinfo {author} {\bibfnamefont {C.}~\bibnamefont
  {Horsman}}, \bibinfo {author} {\bibfnamefont {A.~G.}\ \bibnamefont {Fowler}},
  \bibinfo {author} {\bibfnamefont {S.}~\bibnamefont {Devitt}},\ and\ \bibinfo
  {author} {\bibfnamefont {R.}~\bibnamefont {Van~Meter}},\ }\bibfield  {title}
  {\bibinfo {title} {Surface code quantum computing by lattice surgery},\
  }\href {https://doi.org/10.1088/1367-2630/14/12/123011} {\bibfield  {journal}
  {\bibinfo  {journal} {New J. Phys.}\ }\textbf {\bibinfo {volume} {14}},\
  \bibinfo {pages} {123011} (\bibinfo {year} {2012})},\ \Eprint
  {https://arxiv.org/abs/1111.4022} {arXiv:1111.4022} \BibitemShut {NoStop}%
\bibitem [{\citenamefont {Fowler}\ and\ \citenamefont
  {Gidney}(2018)}]{FowlerGidney2018}%
  \BibitemOpen
  \bibfield  {author} {\bibinfo {author} {\bibfnamefont {A.~G.}\ \bibnamefont
  {Fowler}}\ and\ \bibinfo {author} {\bibfnamefont {C.}~\bibnamefont
  {Gidney}},\ }\bibfield  {title} {\bibinfo {title} {Low overhead quantum
  computation using lattice surgery},\ }\Eprint
  {https://arxiv.org/abs/1808.06709} {arXiv:1808.06709}  (\bibinfo {year}
  {2018})\BibitemShut {NoStop}%
\bibitem [{\citenamefont {Litinski}\ and\ \citenamefont {von
  Oppen}(2018)}]{LitinskiVonOppen2018}%
  \BibitemOpen
  \bibfield  {author} {\bibinfo {author} {\bibfnamefont {D.}~\bibnamefont
  {Litinski}}\ and\ \bibinfo {author} {\bibfnamefont {F.}~\bibnamefont {von
  Oppen}},\ }\bibfield  {title} {\bibinfo {title} {Lattice surgery with a
  twist: Simplifying {Clifford} gates of surface codes},\ }\href
  {https://doi.org/10.22331/q-2018-05-04-62} {\bibfield  {journal} {\bibinfo
  {journal} {Quantum}\ }\textbf {\bibinfo {volume} {2}},\ \bibinfo {pages} {62}
  (\bibinfo {year} {2018})},\ \Eprint {https://arxiv.org/abs/1709.02318}
  {arXiv:1709.02318} \BibitemShut {NoStop}%
\bibitem [{\citenamefont {Litinski}(2019)}]{Litinski2019}%
  \BibitemOpen
  \bibfield  {author} {\bibinfo {author} {\bibfnamefont {D.}~\bibnamefont
  {Litinski}},\ }\bibfield  {title} {\bibinfo {title} {A game of surface codes:
  Large-scale quantum computing with lattice surgery},\ }\href
  {https://doi.org/10.22331/q-2019-03-05-128} {\bibfield  {journal} {\bibinfo
  {journal} {Quantum}\ }\textbf {\bibinfo {volume} {3}},\ \bibinfo {pages}
  {128} (\bibinfo {year} {2019})},\ \Eprint {https://arxiv.org/abs/1808.02892}
  {arXiv:1808.02892} \BibitemShut {NoStop}%
\bibitem [{\citenamefont {{Google Quantum AI}}(2023)}]{Google2023Scaling}%
  \BibitemOpen
  \bibfield  {author} {\bibinfo {author} {\bibnamefont {{Google Quantum AI}}},\
  }\bibfield  {title} {\bibinfo {title} {Suppressing quantum errors by scaling
  a surface code logical qubit},\ }\href
  {https://doi.org/10.1038/s41586-022-05434-1} {\bibfield  {journal} {\bibinfo
  {journal} {Nature}\ }\textbf {\bibinfo {volume} {614}},\ \bibinfo {pages}
  {676} (\bibinfo {year} {2023})},\ \Eprint {https://arxiv.org/abs/2207.06431}
  {arXiv:2207.06431} \BibitemShut {NoStop}%
\bibitem [{\citenamefont {{Google Quantum AI and
  Collaborators}}(2025)}]{Google2025BelowThreshold}%
  \BibitemOpen
  \bibfield  {author} {\bibinfo {author} {\bibnamefont {{Google Quantum AI and
  Collaborators}}},\ }\bibfield  {title} {\bibinfo {title} {Quantum error
  correction below the surface code threshold},\ }\href
  {https://doi.org/10.1038/s41586-024-08449-y} {\bibfield  {journal} {\bibinfo
  {journal} {Nature}\ }\textbf {\bibinfo {volume} {638}},\ \bibinfo {pages}
  {920} (\bibinfo {year} {2025})},\ \Eprint {https://arxiv.org/abs/2408.13687}
  {arXiv:2408.13687} \BibitemShut {NoStop}%
\bibitem [{\citenamefont {Krinner}\ \emph {et~al.}(2022)\citenamefont
  {Krinner}, \citenamefont {Lacroix}, \citenamefont {Remm} \emph
  {et~al.}}]{Krinner2022}%
  \BibitemOpen
  \bibfield  {author} {\bibinfo {author} {\bibfnamefont {S.}~\bibnamefont
  {Krinner}}, \bibinfo {author} {\bibfnamefont {N.}~\bibnamefont {Lacroix}},
  \bibinfo {author} {\bibfnamefont {A.}~\bibnamefont {Remm}}, \emph {et~al.},\
  }\bibfield  {title} {\bibinfo {title} {Realizing repeated quantum error
  correction in a distance-three surface code},\ }\href
  {https://doi.org/10.1038/s41586-022-04566-8} {\bibfield  {journal} {\bibinfo
  {journal} {Nature}\ }\textbf {\bibinfo {volume} {605}},\ \bibinfo {pages}
  {669} (\bibinfo {year} {2022})},\ \Eprint {https://arxiv.org/abs/2112.03708}
  {arXiv:2112.03708} \BibitemShut {NoStop}%
\bibitem [{\citenamefont {Zhao}\ \emph {et~al.}(2022)\citenamefont {Zhao},
  \citenamefont {Ye}, \citenamefont {Huang} \emph {et~al.}}]{Zhao2022}%
  \BibitemOpen
  \bibfield  {author} {\bibinfo {author} {\bibfnamefont {Y.}~\bibnamefont
  {Zhao}}, \bibinfo {author} {\bibfnamefont {Y.}~\bibnamefont {Ye}}, \bibinfo
  {author} {\bibfnamefont {H.-L.}\ \bibnamefont {Huang}}, \emph {et~al.},\
  }\bibfield  {title} {\bibinfo {title} {Realization of an error-correcting
  surface code with superconducting qubits},\ }\href
  {https://doi.org/10.1103/physrevlett.129.030501} {\bibfield  {journal}
  {\bibinfo  {journal} {Phys. Rev. Lett.}\ }\textbf {\bibinfo {volume} {129}},\
  \bibinfo {pages} {030501} (\bibinfo {year} {2022})},\ \Eprint
  {https://arxiv.org/abs/2112.13505} {arXiv:2112.13505} \BibitemShut {NoStop}%
\bibitem [{\citenamefont {Bluvstein}\ \emph {et~al.}(2024)\citenamefont
  {Bluvstein}, \citenamefont {Evered}, \citenamefont {Geim} \emph
  {et~al.}}]{Bluvstein2024}%
  \BibitemOpen
  \bibfield  {author} {\bibinfo {author} {\bibfnamefont {D.}~\bibnamefont
  {Bluvstein}}, \bibinfo {author} {\bibfnamefont {S.~J.}\ \bibnamefont
  {Evered}}, \bibinfo {author} {\bibfnamefont {A.~A.}\ \bibnamefont {Geim}},
  \emph {et~al.},\ }\bibfield  {title} {\bibinfo {title} {Logical quantum
  processor based on reconfigurable atom arrays},\ }\href
  {https://doi.org/10.1038/s41586-023-06927-3} {\bibfield  {journal} {\bibinfo
  {journal} {Nature}\ }\textbf {\bibinfo {volume} {626}},\ \bibinfo {pages}
  {58} (\bibinfo {year} {2024})},\ \Eprint {https://arxiv.org/abs/2312.03982}
  {arXiv:2312.03982} \BibitemShut {NoStop}%
\bibitem [{\citenamefont {Kovalev}\ and\ \citenamefont
  {Pryadko}(2013)}]{KovalevPryadko2013}%
  \BibitemOpen
  \bibfield  {author} {\bibinfo {author} {\bibfnamefont {A.~A.}\ \bibnamefont
  {Kovalev}}\ and\ \bibinfo {author} {\bibfnamefont {L.~P.}\ \bibnamefont
  {Pryadko}},\ }\bibfield  {title} {\bibinfo {title} {Fault tolerance of
  quantum low-density parity check codes with sublinear distance scaling},\
  }\href {https://doi.org/10.1103/physreva.87.020304} {\bibfield  {journal}
  {\bibinfo  {journal} {Phys. Rev. A}\ }\textbf {\bibinfo {volume} {87}},\
  \bibinfo {pages} {020304} (\bibinfo {year} {2013})},\ \Eprint
  {https://arxiv.org/abs/1208.2317} {arXiv:1208.2317} \BibitemShut {NoStop}%
\bibitem [{\citenamefont {Dumer}\ \emph {et~al.}(2015)\citenamefont {Dumer},
  \citenamefont {Kovalev},\ and\ \citenamefont
  {Pryadko}}]{DumerKovalevPryadko2015}%
  \BibitemOpen
  \bibfield  {author} {\bibinfo {author} {\bibfnamefont {I.}~\bibnamefont
  {Dumer}}, \bibinfo {author} {\bibfnamefont {A.~A.}\ \bibnamefont {Kovalev}},\
  and\ \bibinfo {author} {\bibfnamefont {L.~P.}\ \bibnamefont {Pryadko}},\
  }\bibfield  {title} {\bibinfo {title} {Thresholds for correcting errors,
  erasures, and faulty syndrome measurements in degenerate quantum codes},\
  }\href {https://doi.org/10.1103/physrevlett.115.050502} {\bibfield  {journal}
  {\bibinfo  {journal} {Phys. Rev. Lett.}\ }\textbf {\bibinfo {volume} {115}},\
  \bibinfo {pages} {050502} (\bibinfo {year} {2015})},\ \Eprint
  {https://arxiv.org/abs/1412.6172} {arXiv:1412.6172} \BibitemShut {NoStop}%
\bibitem [{\citenamefont {Freedman}\ \emph {et~al.}(2002)\citenamefont
  {Freedman}, \citenamefont {Meyer},\ and\ \citenamefont
  {Luo}}]{FreedmanMeyerLuo2002}%
  \BibitemOpen
  \bibfield  {author} {\bibinfo {author} {\bibfnamefont {M.}~\bibnamefont
  {Freedman}}, \bibinfo {author} {\bibfnamefont {D.}~\bibnamefont {Meyer}},\
  and\ \bibinfo {author} {\bibfnamefont {F.}~\bibnamefont {Luo}},\ }\bibfield
  {title} {\bibinfo {title} {{$Z_2$}-systolic freedom and quantum codes},\ }in\
  \href {https://doi.org/10.1201/9781420035377.ch12} {\emph {\bibinfo
  {booktitle} {Mathematics of Quantum Computation}}}\ (\bibinfo  {publisher}
  {Chapman and Hall/CRC},\ \bibinfo {year} {2002})\BibitemShut {NoStop}%
\bibitem [{\citenamefont {Breuckmann}\ and\ \citenamefont
  {Terhal}(2016)}]{BreuckmannTerhal2016}%
  \BibitemOpen
  \bibfield  {author} {\bibinfo {author} {\bibfnamefont {N.~P.}\ \bibnamefont
  {Breuckmann}}\ and\ \bibinfo {author} {\bibfnamefont {B.~M.}\ \bibnamefont
  {Terhal}},\ }\bibfield  {title} {\bibinfo {title} {Constructions and noise
  threshold of hyperbolic surface codes},\ }\href
  {https://doi.org/10.1109/tit.2016.2555700} {\bibfield  {journal} {\bibinfo
  {journal} {IEEE Trans. Inf. Theory}\ }\textbf {\bibinfo {volume} {62}},\
  \bibinfo {pages} {3731} (\bibinfo {year} {2016})},\ \Eprint
  {https://arxiv.org/abs/1506.04029} {arXiv:1506.04029} \BibitemShut {NoStop}%
\bibitem [{\citenamefont {Breuckmann}\ \emph {et~al.}(2017)\citenamefont
  {Breuckmann}, \citenamefont {Vuillot}, \citenamefont {Campbell},
  \citenamefont {Krishna},\ and\ \citenamefont
  {Terhal}}]{Breuckmann2017SemiHyperbolic}%
  \BibitemOpen
  \bibfield  {author} {\bibinfo {author} {\bibfnamefont {N.~P.}\ \bibnamefont
  {Breuckmann}}, \bibinfo {author} {\bibfnamefont {C.}~\bibnamefont {Vuillot}},
  \bibinfo {author} {\bibfnamefont {E.}~\bibnamefont {Campbell}}, \bibinfo
  {author} {\bibfnamefont {A.}~\bibnamefont {Krishna}},\ and\ \bibinfo {author}
  {\bibfnamefont {B.~M.}\ \bibnamefont {Terhal}},\ }\bibfield  {title}
  {\bibinfo {title} {Hyperbolic and semi-hyperbolic surface codes for quantum
  storage},\ }\href {https://doi.org/10.1088/2058-9565/aa7d3b} {\bibfield
  {journal} {\bibinfo  {journal} {Quantum Sci. Technol.}\ }\textbf {\bibinfo
  {volume} {2}},\ \bibinfo {pages} {035007} (\bibinfo {year} {2017})},\ \Eprint
  {https://arxiv.org/abs/1703.00590} {arXiv:1703.00590} \BibitemShut {NoStop}%
\bibitem [{\citenamefont {Tillich}\ and\ \citenamefont
  {Z{\'e}mor}(2014)}]{TillichZemor2014}%
  \BibitemOpen
  \bibfield  {author} {\bibinfo {author} {\bibfnamefont {J.-P.}\ \bibnamefont
  {Tillich}}\ and\ \bibinfo {author} {\bibfnamefont {G.}~\bibnamefont
  {Z{\'e}mor}},\ }\bibfield  {title} {\bibinfo {title} {Quantum {LDPC} codes
  with positive rate and minimum distance proportional to the square root of
  the blocklength},\ }\href {https://doi.org/10.1109/tit.2013.2292061}
  {\bibfield  {journal} {\bibinfo  {journal} {IEEE Trans. Inf. Theory}\
  }\textbf {\bibinfo {volume} {60}},\ \bibinfo {pages} {1193} (\bibinfo {year}
  {2014})},\ \Eprint {https://arxiv.org/abs/0903.0566} {arXiv:0903.0566}
  \BibitemShut {NoStop}%
\bibitem [{\citenamefont {Panteleev}\ and\ \citenamefont
  {Kalachev}(2022)}]{PanteleevKalachev2022}%
  \BibitemOpen
  \bibfield  {author} {\bibinfo {author} {\bibfnamefont {P.}~\bibnamefont
  {Panteleev}}\ and\ \bibinfo {author} {\bibfnamefont {G.}~\bibnamefont
  {Kalachev}},\ }\bibfield  {title} {\bibinfo {title} {Asymptotically good
  quantum and locally testable classical {LDPC} codes},\ }in\ \href
  {https://doi.org/10.1145/3519935.3520017} {\emph {\bibinfo {booktitle}
  {Proceedings of the 54th Annual ACM SIGACT Symposium on Theory of
  Computing}}}\ (\bibinfo  {publisher} {ACM},\ \bibinfo {year} {2022})\ pp.\
  \bibinfo {pages} {375--388},\ \Eprint {https://arxiv.org/abs/2111.03654}
  {arXiv:2111.03654} \BibitemShut {NoStop}%
\bibitem [{\citenamefont {Leverrier}\ and\ \citenamefont
  {Z{\'e}mor}(2022)}]{LeverrierZemor2022}%
  \BibitemOpen
  \bibfield  {author} {\bibinfo {author} {\bibfnamefont {A.}~\bibnamefont
  {Leverrier}}\ and\ \bibinfo {author} {\bibfnamefont {G.}~\bibnamefont
  {Z{\'e}mor}},\ }\bibfield  {title} {\bibinfo {title} {Quantum {Tanner}
  codes},\ }in\ \href {https://doi.org/10.1109/focs54457.2022.00117} {\emph
  {\bibinfo {booktitle} {2022 IEEE 63rd Annual Symposium on Foundations of
  Computer Science (FOCS)}}}\ (\bibinfo  {publisher} {IEEE},\ \bibinfo {year}
  {2022})\ pp.\ \bibinfo {pages} {872--883},\ \Eprint
  {https://arxiv.org/abs/2202.13641} {arXiv:2202.13641} \BibitemShut {NoStop}%
\bibitem [{\citenamefont {Dinur}\ \emph {et~al.}(2023)\citenamefont {Dinur},
  \citenamefont {Hsieh}, \citenamefont {Lin},\ and\ \citenamefont
  {Vidick}}]{DinurHsiehLinVidick2023}%
  \BibitemOpen
  \bibfield  {author} {\bibinfo {author} {\bibfnamefont {I.}~\bibnamefont
  {Dinur}}, \bibinfo {author} {\bibfnamefont {M.-H.}\ \bibnamefont {Hsieh}},
  \bibinfo {author} {\bibfnamefont {T.-C.}\ \bibnamefont {Lin}},\ and\ \bibinfo
  {author} {\bibfnamefont {T.}~\bibnamefont {Vidick}},\ }\bibfield  {title}
  {\bibinfo {title} {Good quantum {LDPC} codes with linear time decoders},\
  }in\ \href {https://doi.org/10.1145/3564246.3585101} {\emph {\bibinfo
  {booktitle} {Proceedings of the 55th Annual ACM Symposium on Theory of
  Computing}}}\ (\bibinfo  {publisher} {ACM},\ \bibinfo {year} {2023})\ pp.\
  \bibinfo {pages} {905--918},\ \Eprint {https://arxiv.org/abs/2206.07750}
  {arXiv:2206.07750} \BibitemShut {NoStop}%
\bibitem [{\citenamefont {Shor}(1996)}]{Shor1996}%
  \BibitemOpen
  \bibfield  {author} {\bibinfo {author} {\bibfnamefont {P.~W.}\ \bibnamefont
  {Shor}},\ }\bibfield  {title} {\bibinfo {title} {Fault-tolerant quantum
  computation},\ }in\ \href {https://doi.org/10.1109/sfcs.1996.548464} {\emph
  {\bibinfo {booktitle} {Proceedings of 37th Conference on Foundations of
  Computer Science}}}\ (\bibinfo  {publisher} {IEEE},\ \bibinfo {year} {1996})\
  pp.\ \bibinfo {pages} {56--65},\ \Eprint
  {https://arxiv.org/abs/quant-ph/9605011} {arXiv:quant-ph/9605011}
  \BibitemShut {NoStop}%
\bibitem [{\citenamefont {Knill}\ and\ \citenamefont
  {Laflamme}(1996)}]{KnillLaflamme1996}%
  \BibitemOpen
  \bibfield  {author} {\bibinfo {author} {\bibfnamefont {E.}~\bibnamefont
  {Knill}}\ and\ \bibinfo {author} {\bibfnamefont {R.}~\bibnamefont
  {Laflamme}},\ }\bibfield  {title} {\bibinfo {title} {Concatenated quantum
  codes},\ }\Eprint {https://arxiv.org/abs/quant-ph/9608012}
  {arXiv:quant-ph/9608012}  (\bibinfo {year} {1996})\BibitemShut {NoStop}%
\bibitem [{\citenamefont {Aharonov}\ and\ \citenamefont
  {Ben-Or}(2008)}]{AharonovBenOr1999}%
  \BibitemOpen
  \bibfield  {author} {\bibinfo {author} {\bibfnamefont {D.}~\bibnamefont
  {Aharonov}}\ and\ \bibinfo {author} {\bibfnamefont {M.}~\bibnamefont
  {Ben-Or}},\ }\bibfield  {title} {\bibinfo {title} {Fault-tolerant quantum
  computation with constant error rate},\ }\href
  {https://doi.org/10.1137/s0097539799359385} {\bibfield  {journal} {\bibinfo
  {journal} {SIAM J. Comput.}\ }\textbf {\bibinfo {volume} {38}},\ \bibinfo
  {pages} {1207} (\bibinfo {year} {2008})},\ \Eprint
  {https://arxiv.org/abs/quant-ph/9906129} {arXiv:quant-ph/9906129}
  \BibitemShut {NoStop}%
\bibitem [{\citenamefont {Yoshida}\ \emph {et~al.}(2025)\citenamefont
  {Yoshida}, \citenamefont {Tamiya},\ and\ \citenamefont
  {Yamasaki}}]{Yoshida2025}%
  \BibitemOpen
  \bibfield  {author} {\bibinfo {author} {\bibfnamefont {S.}~\bibnamefont
  {Yoshida}}, \bibinfo {author} {\bibfnamefont {S.}~\bibnamefont {Tamiya}},\
  and\ \bibinfo {author} {\bibfnamefont {H.}~\bibnamefont {Yamasaki}},\
  }\bibfield  {title} {\bibinfo {title} {Concatenate codes, save qubits},\
  }\href {https://doi.org/10.1038/s41534-025-01035-8} {\bibfield  {journal}
  {\bibinfo  {journal} {npj Quantum Inf.}\ }\textbf {\bibinfo {volume} {11}},\
  \bibinfo {pages} {88} (\bibinfo {year} {2025})},\ \Eprint
  {https://arxiv.org/abs/2402.09606} {arXiv:2402.09606} \BibitemShut {NoStop}%
\bibitem [{\citenamefont {Bravyi}\ \emph {et~al.}(2024)\citenamefont {Bravyi},
  \citenamefont {Cross}, \citenamefont {Gambetta}, \citenamefont {Maslov},
  \citenamefont {Rall},\ and\ \citenamefont {Yoder}}]{Bravyi2024BB}%
  \BibitemOpen
  \bibfield  {author} {\bibinfo {author} {\bibfnamefont {S.}~\bibnamefont
  {Bravyi}}, \bibinfo {author} {\bibfnamefont {A.~W.}\ \bibnamefont {Cross}},
  \bibinfo {author} {\bibfnamefont {J.~M.}\ \bibnamefont {Gambetta}}, \bibinfo
  {author} {\bibfnamefont {D.}~\bibnamefont {Maslov}}, \bibinfo {author}
  {\bibfnamefont {P.}~\bibnamefont {Rall}},\ and\ \bibinfo {author}
  {\bibfnamefont {T.~J.}\ \bibnamefont {Yoder}},\ }\bibfield  {title} {\bibinfo
  {title} {High-threshold and low-overhead fault-tolerant quantum memory},\
  }\href {https://doi.org/10.1038/s41586-024-07107-7} {\bibfield  {journal}
  {\bibinfo  {journal} {Nature}\ }\textbf {\bibinfo {volume} {627}},\ \bibinfo
  {pages} {778} (\bibinfo {year} {2024})},\ \Eprint
  {https://arxiv.org/abs/2308.07915} {arXiv:2308.07915} \BibitemShut {NoStop}%
\bibitem [{\citenamefont {Gidney}\ \emph {et~al.}(2025)\citenamefont {Gidney},
  \citenamefont {Newman}, \citenamefont {Brooks},\ and\ \citenamefont
  {Jones}}]{Gidney2025Yoked}%
  \BibitemOpen
  \bibfield  {author} {\bibinfo {author} {\bibfnamefont {C.}~\bibnamefont
  {Gidney}}, \bibinfo {author} {\bibfnamefont {M.}~\bibnamefont {Newman}},
  \bibinfo {author} {\bibfnamefont {P.}~\bibnamefont {Brooks}},\ and\ \bibinfo
  {author} {\bibfnamefont {C.}~\bibnamefont {Jones}},\ }\bibfield  {title}
  {\bibinfo {title} {Yoked surface codes},\ }\href
  {https://doi.org/10.1038/s41467-025-59714-1} {\bibfield  {journal} {\bibinfo
  {journal} {Nat. Commun.}\ }\textbf {\bibinfo {volume} {16}},\ \bibinfo
  {pages} {4498} (\bibinfo {year} {2025})},\ \Eprint
  {https://arxiv.org/abs/2312.04522} {arXiv:2312.04522} \BibitemShut {NoStop}%
\bibitem [{\citenamefont {Pattison}\ \emph {et~al.}(2025)\citenamefont
  {Pattison}, \citenamefont {Krishna},\ and\ \citenamefont
  {Preskill}}]{Pattison2025}%
  \BibitemOpen
  \bibfield  {author} {\bibinfo {author} {\bibfnamefont {C.~A.}\ \bibnamefont
  {Pattison}}, \bibinfo {author} {\bibfnamefont {A.}~\bibnamefont {Krishna}},\
  and\ \bibinfo {author} {\bibfnamefont {J.}~\bibnamefont {Preskill}},\
  }\bibfield  {title} {\bibinfo {title} {Hierarchical memories: Simulating
  quantum {LDPC} codes with local gates},\ }\href
  {https://doi.org/10.22331/q-2025-05-05-1728} {\bibfield  {journal} {\bibinfo
  {journal} {Quantum}\ }\textbf {\bibinfo {volume} {9}},\ \bibinfo {pages}
  {1728} (\bibinfo {year} {2025})},\ \Eprint {https://arxiv.org/abs/2303.04798}
  {arXiv:2303.04798} \BibitemShut {NoStop}%
\bibitem [{\citenamefont {Vaidman}\ \emph {et~al.}(1996)\citenamefont
  {Vaidman}, \citenamefont {Goldenberg},\ and\ \citenamefont
  {Wiesner}}]{Vaidman1996}%
  \BibitemOpen
  \bibfield  {author} {\bibinfo {author} {\bibfnamefont {L.}~\bibnamefont
  {Vaidman}}, \bibinfo {author} {\bibfnamefont {L.}~\bibnamefont
  {Goldenberg}},\ and\ \bibinfo {author} {\bibfnamefont {S.}~\bibnamefont
  {Wiesner}},\ }\bibfield  {title} {\bibinfo {title} {Error prevention scheme
  with four particles},\ }\href {https://doi.org/10.1103/PhysRevA.54.R1745}
  {\bibfield  {journal} {\bibinfo  {journal} {Phys. Rev. A}\ }\textbf {\bibinfo
  {volume} {54}},\ \bibinfo {pages} {R1745} (\bibinfo {year} {1996})},\ \Eprint
  {https://arxiv.org/abs/quant-ph/9603031} {arXiv:quant-ph/9603031}
  \BibitemShut {NoStop}%
\bibitem [{\citenamefont {Grassl}\ \emph {et~al.}(1997)\citenamefont {Grassl},
  \citenamefont {Beth},\ and\ \citenamefont
  {Pellizzari}}]{GrasslBethPellizzari1997}%
  \BibitemOpen
  \bibfield  {author} {\bibinfo {author} {\bibfnamefont {M.}~\bibnamefont
  {Grassl}}, \bibinfo {author} {\bibfnamefont {T.}~\bibnamefont {Beth}},\ and\
  \bibinfo {author} {\bibfnamefont {T.}~\bibnamefont {Pellizzari}},\ }\bibfield
   {title} {\bibinfo {title} {Codes for the quantum erasure channel},\ }\href
  {https://doi.org/10.1103/PhysRevA.56.33} {\bibfield  {journal} {\bibinfo
  {journal} {Phys. Rev. A}\ }\textbf {\bibinfo {volume} {56}},\ \bibinfo
  {pages} {33} (\bibinfo {year} {1997})},\ \Eprint
  {https://arxiv.org/abs/quant-ph/9610042} {arXiv:quant-ph/9610042}
  \BibitemShut {NoStop}%
\bibitem [{\citenamefont {Knill}(2005)}]{Knill2005}%
  \BibitemOpen
  \bibfield  {author} {\bibinfo {author} {\bibfnamefont {E.}~\bibnamefont
  {Knill}},\ }\bibfield  {title} {\bibinfo {title} {Quantum computing with
  realistically noisy devices},\ }\href {https://doi.org/10.1038/nature03350}
  {\bibfield  {journal} {\bibinfo  {journal} {Nature}\ }\textbf {\bibinfo
  {volume} {434}},\ \bibinfo {pages} {39} (\bibinfo {year} {2005})},\ \Eprint
  {https://arxiv.org/abs/quant-ph/0410199} {arXiv:quant-ph/0410199}
  \BibitemShut {NoStop}%
\bibitem [{\citenamefont {Gidney}(2021)}]{Gidney2021Stim}%
  \BibitemOpen
  \bibfield  {author} {\bibinfo {author} {\bibfnamefont {C.}~\bibnamefont
  {Gidney}},\ }\bibfield  {title} {\bibinfo {title} {Stim: A fast stabilizer
  circuit simulator},\ }\href {https://doi.org/10.22331/q-2021-07-06-497}
  {\bibfield  {journal} {\bibinfo  {journal} {Quantum}\ }\textbf {\bibinfo
  {volume} {5}},\ \bibinfo {pages} {497} (\bibinfo {year} {2021})},\ \Eprint
  {https://arxiv.org/abs/2103.02202} {arXiv:2103.02202} \BibitemShut {NoStop}%
\bibitem [{\citenamefont {Panteleev}\ and\ \citenamefont
  {Kalachev}(2021)}]{PanteleevKalachev2021}%
  \BibitemOpen
  \bibfield  {author} {\bibinfo {author} {\bibfnamefont {P.}~\bibnamefont
  {Panteleev}}\ and\ \bibinfo {author} {\bibfnamefont {G.}~\bibnamefont
  {Kalachev}},\ }\bibfield  {title} {\bibinfo {title} {Degenerate quantum
  {LDPC} codes with good finite length performance},\ }\href
  {https://doi.org/10.22331/q-2021-11-22-585} {\bibfield  {journal} {\bibinfo
  {journal} {Quantum}\ }\textbf {\bibinfo {volume} {5}},\ \bibinfo {pages}
  {585} (\bibinfo {year} {2021})},\ \Eprint {https://arxiv.org/abs/1904.02703}
  {arXiv:1904.02703} \BibitemShut {NoStop}%
\bibitem [{\citenamefont {Roffe}\ \emph {et~al.}(2020)\citenamefont {Roffe},
  \citenamefont {White}, \citenamefont {Burton},\ and\ \citenamefont
  {Campbell}}]{Roffe2020}%
  \BibitemOpen
  \bibfield  {author} {\bibinfo {author} {\bibfnamefont {J.}~\bibnamefont
  {Roffe}}, \bibinfo {author} {\bibfnamefont {D.~R.}\ \bibnamefont {White}},
  \bibinfo {author} {\bibfnamefont {S.}~\bibnamefont {Burton}},\ and\ \bibinfo
  {author} {\bibfnamefont {E.}~\bibnamefont {Campbell}},\ }\bibfield  {title}
  {\bibinfo {title} {Decoding across the quantum low-density parity-check code
  landscape},\ }\href {https://doi.org/10.1103/physrevresearch.2.043423}
  {\bibfield  {journal} {\bibinfo  {journal} {Phys. Rev. Res.}\ }\textbf
  {\bibinfo {volume} {2}},\ \bibinfo {pages} {043423} (\bibinfo {year}
  {2020})},\ \Eprint {https://arxiv.org/abs/2005.07016} {arXiv:2005.07016}
  \BibitemShut {NoStop}%
\bibitem [{\citenamefont {Aghababaie~Beni}\ \emph {et~al.}(2025)\citenamefont
  {Aghababaie~Beni}, \citenamefont {Higgott},\ and\ \citenamefont
  {Shutty}}]{Tesseract2025}%
  \BibitemOpen
  \bibfield  {author} {\bibinfo {author} {\bibfnamefont {L.}~\bibnamefont
  {Aghababaie~Beni}}, \bibinfo {author} {\bibfnamefont {O.}~\bibnamefont
  {Higgott}},\ and\ \bibinfo {author} {\bibfnamefont {N.}~\bibnamefont
  {Shutty}},\ }\bibfield  {title} {\bibinfo {title} {Tesseract: a search-based
  decoder for quantum error correction},\ }\Eprint
  {https://arxiv.org/abs/2503.10988} {arXiv:2503.10988}  (\bibinfo {year}
  {2025})\BibitemShut {NoStop}%
\bibitem [{\citenamefont {Higgott}\ and\ \citenamefont
  {Gidney}(2025)}]{Higgott2025SparseBlossom}%
  \BibitemOpen
  \bibfield  {author} {\bibinfo {author} {\bibfnamefont {O.}~\bibnamefont
  {Higgott}}\ and\ \bibinfo {author} {\bibfnamefont {C.}~\bibnamefont
  {Gidney}},\ }\bibfield  {title} {\bibinfo {title} {Sparse blossom: correcting
  a million errors per core second with minimum-weight matching},\ }\href
  {https://doi.org/10.22331/q-2025-01-20-1600} {\bibfield  {journal} {\bibinfo
  {journal} {Quantum}\ }\textbf {\bibinfo {volume} {9}},\ \bibinfo {pages}
  {1600} (\bibinfo {year} {2025})},\ \Eprint {https://arxiv.org/abs/2303.15933}
  {arXiv:2303.15933} \BibitemShut {NoStop}%
\bibitem [{\citenamefont {Fowler}(2013)}]{Fowler2013Correlated}%
  \BibitemOpen
  \bibfield  {author} {\bibinfo {author} {\bibfnamefont {A.~G.}\ \bibnamefont
  {Fowler}},\ }\bibfield  {title} {\bibinfo {title} {Optimal complexity
  correction of correlated errors in the surface code},\ }\Eprint
  {https://arxiv.org/abs/1310.0863} {arXiv:1310.0863}  (\bibinfo {year}
  {2013})\BibitemShut {NoStop}%
\bibitem [{\citenamefont {Gidney}\ \emph {et~al.}(2021)\citenamefont {Gidney},
  \citenamefont {Newman}, \citenamefont {Fowler},\ and\ \citenamefont
  {Broughton}}]{GidneyNewmanFowlerBroughton2021}%
  \BibitemOpen
  \bibfield  {author} {\bibinfo {author} {\bibfnamefont {C.}~\bibnamefont
  {Gidney}}, \bibinfo {author} {\bibfnamefont {M.}~\bibnamefont {Newman}},
  \bibinfo {author} {\bibfnamefont {A.}~\bibnamefont {Fowler}},\ and\ \bibinfo
  {author} {\bibfnamefont {M.}~\bibnamefont {Broughton}},\ }\bibfield  {title}
  {\bibinfo {title} {A fault-tolerant honeycomb memory},\ }\href
  {https://doi.org/10.22331/q-2021-12-20-605} {\bibfield  {journal} {\bibinfo
  {journal} {Quantum}\ }\textbf {\bibinfo {volume} {5}},\ \bibinfo {pages}
  {605} (\bibinfo {year} {2021})},\ \Eprint {https://arxiv.org/abs/2108.10457}
  {arXiv:2108.10457} \BibitemShut {NoStop}%
\bibitem [{\citenamefont {Bombin}\ and\ \citenamefont
  {Martin-Delgado}(2007)}]{BombinMartinDelgado2007}%
  \BibitemOpen
  \bibfield  {author} {\bibinfo {author} {\bibfnamefont {H.}~\bibnamefont
  {Bombin}}\ and\ \bibinfo {author} {\bibfnamefont {M.~A.}\ \bibnamefont
  {Martin-Delgado}},\ }\bibfield  {title} {\bibinfo {title} {Optimal resources
  for topological two-dimensional stabilizer codes: Comparative study},\ }\href
  {https://doi.org/10.1103/PhysRevA.76.012305} {\bibfield  {journal} {\bibinfo
  {journal} {Phys. Rev. A}\ }\textbf {\bibinfo {volume} {76}},\ \bibinfo
  {pages} {012305} (\bibinfo {year} {2007})},\ \Eprint
  {https://arxiv.org/abs/quant-ph/0703272} {arXiv:quant-ph/0703272}
  \BibitemShut {NoStop}%
\bibitem [{\citenamefont {Gottesman}(2016)}]{Gottesman2016Small}%
  \BibitemOpen
  \bibfield  {author} {\bibinfo {author} {\bibfnamefont {D.}~\bibnamefont
  {Gottesman}},\ }\bibfield  {title} {\bibinfo {title} {Quantum fault tolerance
  in small experiments},\ }\Eprint {https://arxiv.org/abs/1610.03507}
  {arXiv:1610.03507}  (\bibinfo {year} {2016})\BibitemShut {NoStop}%
\bibitem [{\citenamefont {Aliferis}\ \emph {et~al.}(2008)\citenamefont
  {Aliferis}, \citenamefont {Gottesman},\ and\ \citenamefont
  {Preskill}}]{AliferisGottesmanPreskill2008}%
  \BibitemOpen
  \bibfield  {author} {\bibinfo {author} {\bibfnamefont {P.}~\bibnamefont
  {Aliferis}}, \bibinfo {author} {\bibfnamefont {D.}~\bibnamefont
  {Gottesman}},\ and\ \bibinfo {author} {\bibfnamefont {J.}~\bibnamefont
  {Preskill}},\ }\bibfield  {title} {\bibinfo {title} {Accuracy threshold for
  postselected quantum computation},\ }\href
  {https://doi.org/10.26421/qic8.3-4-1} {\bibfield  {journal} {\bibinfo
  {journal} {Quant. Inf. Comput.}\ }\textbf {\bibinfo {volume} {8}},\ \bibinfo
  {pages} {181} (\bibinfo {year} {2008})},\ \Eprint
  {https://arxiv.org/abs/quant-ph/0703264} {arXiv:quant-ph/0703264}
  \BibitemShut {NoStop}%
\bibitem [{\citenamefont {Steane}(1997)}]{Steane1997}%
  \BibitemOpen
  \bibfield  {author} {\bibinfo {author} {\bibfnamefont {A.~M.}\ \bibnamefont
  {Steane}},\ }\bibfield  {title} {\bibinfo {title} {Active stabilization,
  quantum computation, and quantum state synthesis},\ }\href
  {https://doi.org/10.1103/PhysRevLett.78.2252} {\bibfield  {journal} {\bibinfo
   {journal} {Phys. Rev. Lett.}\ }\textbf {\bibinfo {volume} {78}},\ \bibinfo
  {pages} {2252} (\bibinfo {year} {1997})},\ \Eprint
  {https://arxiv.org/abs/quant-ph/9611027} {arXiv:quant-ph/9611027}
  \BibitemShut {NoStop}%
\bibitem [{\citenamefont {DiVincenzo}\ and\ \citenamefont
  {Aliferis}(2007)}]{DiVincenzoAliferis2007}%
  \BibitemOpen
  \bibfield  {author} {\bibinfo {author} {\bibfnamefont {D.~P.}\ \bibnamefont
  {DiVincenzo}}\ and\ \bibinfo {author} {\bibfnamefont {P.}~\bibnamefont
  {Aliferis}},\ }\bibfield  {title} {\bibinfo {title} {Effective fault-tolerant
  quantum computation with slow measurements},\ }\href
  {https://doi.org/10.1103/physrevlett.98.020501} {\bibfield  {journal}
  {\bibinfo  {journal} {Phys. Rev. Lett.}\ }\textbf {\bibinfo {volume} {98}},\
  \bibinfo {pages} {020501} (\bibinfo {year} {2007})},\ \Eprint
  {https://arxiv.org/abs/quant-ph/0607047} {arXiv:quant-ph/0607047}
  \BibitemShut {NoStop}%
\bibitem [{\citenamefont {Bao}\ \emph {et~al.}(2024)\citenamefont {Bao},
  \citenamefont {Xu}, \citenamefont {Song} \emph {et~al.}}]{Bao2024GHZ}%
  \BibitemOpen
  \bibfield  {author} {\bibinfo {author} {\bibfnamefont {Z.}~\bibnamefont
  {Bao}}, \bibinfo {author} {\bibfnamefont {S.}~\bibnamefont {Xu}}, \bibinfo
  {author} {\bibfnamefont {Z.}~\bibnamefont {Song}}, \emph {et~al.},\
  }\bibfield  {title} {\bibinfo {title} {Creating and controlling global
  {Greenberger-Horne-Zeilinger} entanglement on quantum processors},\ }\href
  {https://doi.org/10.1038/s41467-024-53140-5} {\bibfield  {journal} {\bibinfo
  {journal} {Nat. Commun.}\ }\textbf {\bibinfo {volume} {15}},\ \bibinfo
  {pages} {8823} (\bibinfo {year} {2024})},\ \Eprint
  {https://arxiv.org/abs/2401.08284} {arXiv:2401.08284} \BibitemShut {NoStop}%
\bibitem [{\citenamefont {Javadi-Abhari}\ \emph {et~al.}(2025)\citenamefont
  {Javadi-Abhari}, \citenamefont {Martiel}, \citenamefont {Seif}, \citenamefont
  {Takita},\ and\ \citenamefont {Wei}}]{IBM2025GHZ}%
  \BibitemOpen
  \bibfield  {author} {\bibinfo {author} {\bibfnamefont {A.}~\bibnamefont
  {Javadi-Abhari}}, \bibinfo {author} {\bibfnamefont {S.}~\bibnamefont
  {Martiel}}, \bibinfo {author} {\bibfnamefont {A.}~\bibnamefont {Seif}},
  \bibinfo {author} {\bibfnamefont {M.}~\bibnamefont {Takita}},\ and\ \bibinfo
  {author} {\bibfnamefont {K.~X.}\ \bibnamefont {Wei}},\ }\bibfield  {title}
  {\bibinfo {title} {Big cats: entanglement in 120 qubits and beyond},\
  }\Eprint {https://arxiv.org/abs/2510.09520} {arXiv:2510.09520}  (\bibinfo
  {year} {2025})\BibitemShut {NoStop}%
\bibitem [{\citenamefont {B{\"{a}}umer}\ \emph {et~al.}(2024)\citenamefont
  {B{\"{a}}umer}, \citenamefont {Tripathi}, \citenamefont {Wang}, \citenamefont
  {Rall}, \citenamefont {Chen}, \citenamefont {Majumder}, \citenamefont
  {Seif},\ and\ \citenamefont {Minev}}]{Baumer2024Dynamic}%
  \BibitemOpen
  \bibfield  {author} {\bibinfo {author} {\bibfnamefont {E.}~\bibnamefont
  {B{\"{a}}umer}}, \bibinfo {author} {\bibfnamefont {V.}~\bibnamefont
  {Tripathi}}, \bibinfo {author} {\bibfnamefont {D.~S.}\ \bibnamefont {Wang}},
  \bibinfo {author} {\bibfnamefont {P.}~\bibnamefont {Rall}}, \bibinfo {author}
  {\bibfnamefont {E.~H.}\ \bibnamefont {Chen}}, \bibinfo {author}
  {\bibfnamefont {S.}~\bibnamefont {Majumder}}, \bibinfo {author}
  {\bibfnamefont {A.}~\bibnamefont {Seif}},\ and\ \bibinfo {author}
  {\bibfnamefont {Z.~K.}\ \bibnamefont {Minev}},\ }\bibfield  {title} {\bibinfo
  {title} {Efficient long-range entanglement using dynamic circuits},\ }\href
  {https://doi.org/10.1103/prxquantum.5.030339} {\bibfield  {journal} {\bibinfo
   {journal} {PRX Quantum}\ }\textbf {\bibinfo {volume} {5}},\ \bibinfo {pages}
  {030339} (\bibinfo {year} {2024})},\ \Eprint
  {https://arxiv.org/abs/2308.13065} {arXiv:2308.13065} \BibitemShut {NoStop}%
\bibitem [{\citenamefont {Pryadko}\ and\ \citenamefont
  {Zeng}(2024)}]{distm4ri}%
  \BibitemOpen
  \bibfield  {author} {\bibinfo {author} {\bibfnamefont {L.~P.}\ \bibnamefont
  {Pryadko}}\ and\ \bibinfo {author} {\bibfnamefont {W.}~\bibnamefont {Zeng}},\
  }\href {https://github.com/QEC-pages/dist-m4ri} {\bibinfo {title} {dist-m4ri:
  distance of a classical or quantum {CSS} code}} (\bibinfo {year} {2024}),\
  \bibinfo {note} {version 0.9.0}\BibitemShut {NoStop}%
\bibitem [{\citenamefont {Higgott}(2022)}]{Higgott2022PyMatching}%
  \BibitemOpen
  \bibfield  {author} {\bibinfo {author} {\bibfnamefont {O.}~\bibnamefont
  {Higgott}},\ }\bibfield  {title} {\bibinfo {title} {{PyMatching}: A {Python}
  package for decoding quantum codes with minimum-weight perfect matching},\
  }\href {https://doi.org/10.1145/3505637} {\bibfield  {journal} {\bibinfo
  {journal} {ACM Trans. Quantum Comput.}\ }\textbf {\bibinfo {volume} {3}},\
  \bibinfo {pages} {16} (\bibinfo {year} {2022})},\ \Eprint
  {https://arxiv.org/abs/2105.13082} {arXiv:2105.13082} \BibitemShut {NoStop}%
\bibitem [{\citenamefont {Sivak}\ \emph {et~al.}(2026)\citenamefont {Sivak},
  \citenamefont {Morvan}, \citenamefont {Broughton} \emph
  {et~al.}}]{sivak2025reinforcement}%
  \BibitemOpen
  \bibfield  {author} {\bibinfo {author} {\bibfnamefont {V.}~\bibnamefont
  {Sivak}}, \bibinfo {author} {\bibfnamefont {A.}~\bibnamefont {Morvan}},
  \bibinfo {author} {\bibfnamefont {M.}~\bibnamefont {Broughton}}, \emph
  {et~al.},\ }\bibfield  {title} {\bibinfo {title} {Reinforcement learning
  control of quantum error correction},\ }\href
  {https://doi.org/10.1038/s41586-026-10759-2} {\bibfield  {journal} {\bibinfo
  {journal} {Nature}\ }\textbf {\bibinfo {volume} {655}},\ \bibinfo {pages}
  {879} (\bibinfo {year} {2026})},\ \Eprint {https://arxiv.org/abs/2511.08493}
  {arXiv:2511.08493} \BibitemShut {NoStop}%
\bibitem [{\citenamefont {Miao}\ \emph {et~al.}(2023)\citenamefont {Miao},
  \citenamefont {McEwen}, \citenamefont {Atalaya} \emph
  {et~al.}}]{Miao2023Leakage}%
  \BibitemOpen
  \bibfield  {author} {\bibinfo {author} {\bibfnamefont {K.~C.}\ \bibnamefont
  {Miao}}, \bibinfo {author} {\bibfnamefont {M.}~\bibnamefont {McEwen}},
  \bibinfo {author} {\bibfnamefont {J.}~\bibnamefont {Atalaya}}, \emph
  {et~al.},\ }\bibfield  {title} {\bibinfo {title} {Overcoming leakage in
  quantum error correction},\ }\href
  {https://doi.org/10.1038/s41567-023-02226-w} {\bibfield  {journal} {\bibinfo
  {journal} {Nat. Phys.}\ }\textbf {\bibinfo {volume} {19}},\ \bibinfo {pages}
  {1780} (\bibinfo {year} {2023})},\ \Eprint {https://arxiv.org/abs/2211.04728}
  {arXiv:2211.04728} \BibitemShut {NoStop}%
\bibitem [{\citenamefont {{Google Quantum AI}}(2021)}]{Google2021Exponential}%
  \BibitemOpen
  \bibfield  {author} {\bibinfo {author} {\bibnamefont {{Google Quantum AI}}},\
  }\bibfield  {title} {\bibinfo {title} {Exponential suppression of bit or
  phase errors with cyclic error correction},\ }\href
  {https://doi.org/10.1038/s41586-021-03588-y} {\bibfield  {journal} {\bibinfo
  {journal} {Nature}\ }\textbf {\bibinfo {volume} {595}},\ \bibinfo {pages}
  {383} (\bibinfo {year} {2021})},\ \Eprint {https://arxiv.org/abs/2102.06132}
  {arXiv:2102.06132} \BibitemShut {NoStop}%
\bibitem [{\citenamefont {Bonilla~Ataides}\ \emph {et~al.}(2021)\citenamefont
  {Bonilla~Ataides}, \citenamefont {Tuckett}, \citenamefont {Bartlett},
  \citenamefont {Flammia},\ and\ \citenamefont
  {Brown}}]{BonillaAtaides2021XZZX}%
  \BibitemOpen
  \bibfield  {author} {\bibinfo {author} {\bibfnamefont {J.~P.}\ \bibnamefont
  {Bonilla~Ataides}}, \bibinfo {author} {\bibfnamefont {D.~K.}\ \bibnamefont
  {Tuckett}}, \bibinfo {author} {\bibfnamefont {S.~D.}\ \bibnamefont
  {Bartlett}}, \bibinfo {author} {\bibfnamefont {S.~T.}\ \bibnamefont
  {Flammia}},\ and\ \bibinfo {author} {\bibfnamefont {B.~J.}\ \bibnamefont
  {Brown}},\ }\bibfield  {title} {\bibinfo {title} {The {XZZX} surface code},\
  }\href {https://doi.org/10.1038/s41467-021-22274-1} {\bibfield  {journal}
  {\bibinfo  {journal} {Nat. Commun.}\ }\textbf {\bibinfo {volume} {12}},\
  \bibinfo {pages} {2172} (\bibinfo {year} {2021})},\ \Eprint
  {https://arxiv.org/abs/2009.07851} {arXiv:2009.07851} \BibitemShut {NoStop}%
\bibitem [{\citenamefont {Gidney}(2022)}]{Gidney2022Stability}%
  \BibitemOpen
  \bibfield  {author} {\bibinfo {author} {\bibfnamefont {C.}~\bibnamefont
  {Gidney}},\ }\bibfield  {title} {\bibinfo {title} {Stability experiments: The
  overlooked dual of memory experiments},\ }\href
  {https://doi.org/10.22331/q-2022-08-24-786} {\bibfield  {journal} {\bibinfo
  {journal} {Quantum}\ }\textbf {\bibinfo {volume} {6}},\ \bibinfo {pages}
  {786} (\bibinfo {year} {2022})},\ \Eprint {https://arxiv.org/abs/2204.13834}
  {arXiv:2204.13834} \BibitemShut {NoStop}%
\bibitem [{\citenamefont {Grassl}\ and\ \citenamefont
  {Roetteler}(2013)}]{GrasslRoetteler2013}%
  \BibitemOpen
  \bibfield  {author} {\bibinfo {author} {\bibfnamefont {M.}~\bibnamefont
  {Grassl}}\ and\ \bibinfo {author} {\bibfnamefont {M.}~\bibnamefont
  {Roetteler}},\ }\bibfield  {title} {\bibinfo {title} {Leveraging
  automorphisms of quantum codes for fault-tolerant quantum computation},\ }in\
  \href {https://doi.org/10.1109/isit.2013.6620283} {\emph {\bibinfo
  {booktitle} {2013 IEEE International Symposium on Information Theory}}}\
  (\bibinfo  {publisher} {IEEE},\ \bibinfo {year} {2013})\ pp.\ \bibinfo
  {pages} {534--538},\ \Eprint {https://arxiv.org/abs/1302.1035}
  {arXiv:1302.1035} \BibitemShut {NoStop}%
\bibitem [{\citenamefont {Breuckmann}\ and\ \citenamefont
  {Burton}(2024)}]{BreuckmannBurton2024}%
  \BibitemOpen
  \bibfield  {author} {\bibinfo {author} {\bibfnamefont {N.~P.}\ \bibnamefont
  {Breuckmann}}\ and\ \bibinfo {author} {\bibfnamefont {S.}~\bibnamefont
  {Burton}},\ }\bibfield  {title} {\bibinfo {title} {Fold-transversal
  {Clifford} gates for quantum codes},\ }\href
  {https://doi.org/10.22331/q-2024-06-13-1372} {\bibfield  {journal} {\bibinfo
  {journal} {Quantum}\ }\textbf {\bibinfo {volume} {8}},\ \bibinfo {pages}
  {1372} (\bibinfo {year} {2024})},\ \Eprint {https://arxiv.org/abs/2202.06647}
  {arXiv:2202.06647} \BibitemShut {NoStop}%
\bibitem [{\citenamefont {Sayginel}\ \emph {et~al.}(2025)\citenamefont
  {Sayginel}, \citenamefont {Koutsioumpas}, \citenamefont {Webster},
  \citenamefont {Rajput},\ and\ \citenamefont {Browne}}]{Sayginel2025}%
  \BibitemOpen
  \bibfield  {author} {\bibinfo {author} {\bibfnamefont {H.}~\bibnamefont
  {Sayginel}}, \bibinfo {author} {\bibfnamefont {S.}~\bibnamefont
  {Koutsioumpas}}, \bibinfo {author} {\bibfnamefont {M.}~\bibnamefont
  {Webster}}, \bibinfo {author} {\bibfnamefont {A.}~\bibnamefont {Rajput}},\
  and\ \bibinfo {author} {\bibfnamefont {D.~E.}\ \bibnamefont {Browne}},\
  }\bibfield  {title} {\bibinfo {title} {Fault-tolerant logical {Clifford}
  gates from code automorphisms},\ }\href {https://doi.org/10.1103/vf7v-cpq9}
  {\bibfield  {journal} {\bibinfo  {journal} {PRX Quantum}\ }\textbf {\bibinfo
  {volume} {6}},\ \bibinfo {pages} {030343} (\bibinfo {year} {2025})},\ \Eprint
  {https://arxiv.org/abs/2409.18175} {arXiv:2409.18175} \BibitemShut {NoStop}%
\end{thebibliography}%

\appendix
\onecolumngrid

\section{Circuit distance and verification depth \boldmath$\rv$}
\label{app:circuit_distance}

We compute the circuit distance with Stim's \texttt{shortest\_graphlike\_error}~\cite{Gidney2021Stim} on the CZ-compiled memory circuits with SI1000 fault locations~\cite{GidneyNewmanFowlerBroughton2021}, for both memory bases ($ZZ$ and $XX$) and for one to three outer-code periods ($\ro=1$--$3$); the result is the same in all six cases.
The graphlike search ignores fault mechanisms that flip more than two detectors and therefore returns an upper bound. An exhaustive connected-cluster search with dist-m4ri~\cite{distm4ri} on the full detector error model, including such faults, confirms every entry of Table~\ref{tab:circuit_distance} up to $7$ exactly; the larger entries are upper bounds.
Table~\ref{tab:circuit_distance} lists representative results.
Every computed value with $\rv \geq 2$ equals $\min(2t,\rv+3)$; at $\rv=1$ the distance is $4$ for even $t$ and $3$ for odd $t$.
We checked the saturation point $\rv=2t-3$ for all $t$ from $3$ to $10$: at $t=10$, for example, $d_{\rm circ}=19$ at $\rv=16$ and $20$ at $\rv=17$.

Below saturation, the minimum-weight logical error sits at the corridor intersection. It consists of a time-like chain of $\rv$ faults that repeats through the verification rounds at a single corridor site, together with three two-qubit-gate faults that close the chain onto a logical representative; each additional verification round lengthens the chain by one. At saturation, the minimum-weight error becomes a spatial logical string of weight $t$ in each of two patches.

\begin{table}[h]
\centering
\footnotesize
\begin{tabular}{@{}c c c c c c c c c c c@{}}
\toprule
& & \multicolumn{9}{c}{$\rv$} \\
\cmidrule(l){3-11}
$t$ & $2t$ & \makebox[2.2em]{$1$} & \makebox[2.2em]{$2$} & \makebox[2.2em]{$3$} & \makebox[2.2em]{$4$} & \makebox[2.2em]{$5$} & \makebox[2.2em]{$6$} & \makebox[2.2em]{$7$} & \makebox[2.2em]{$8$} & \makebox[2.2em]{$9$} \\
\midrule
$2$ & $4$ & $4$ & $4$ & $4$ & $4$ & $4$ & $4$ & $4$ & $4$ & $4$ \\
$3$ & $6$ & $3$ & $5$ & $6$ & $6$ & $6$ & $6$ & $6$ & $6$ & $6$ \\
$4$ & $8$ & $4$ & $5$ & $6$ & $7$ & $\le 8$ & $\le 8$ & $\le 8$ & $\le 8$ & $\le 8$ \\
$5$ & $10$ & $3$ & $5$ & $6$ & $7$ & $\le 8$ & $\le 9$ & $\le 10$ & $\le 10$ & $\le 10$ \\
$6$ & $12$ & $4$ & $5$ & $6$ & $7$ & $\le 8$ & $\le 9$ & $\le 10$ & $\le 11$ & $\le 12$ \\
\bottomrule
\end{tabular}
\caption{Circuit distance $d_{\rm circ}$ of the stitched memory circuit versus verification depth $\rv$, from Stim's graphlike distance search under SI1000 fault locations; the values are the same for the $ZZ$ and $XX$ memory circuits and for $\ro=1$--$3$. Entries up to $7$ are exact (dist-m4ri); larger entries are upper bounds. For $\rv\geq2$ the entries follow $\min(2t,\rv+3)$.}
\label{tab:circuit_distance}
\end{table}

\section{Markov chain model for logical errors}
\label{app:markov-chain}
We approximate the error dynamics for the two logical qubits with a stationary continuous-time four-state Markov chain, with the states labeled $\{00,01,10,11\}$, where a non-zero bit corresponds to a logical error in the corresponding channel.    We ignore any $Y$ errors (which are expected to be strongly suppressed), and consider a symmetric model where two identical errors cancel each other (or, equivalently, the rates $0\to1$ and $1\to 0$ are equal, as would be expected for a Pauli channel).  Denoting the three error rates as $\mu_{01}$, $\mu_{10}$, $\mu_{11}$, and their sum as $\mu_{\rm tot}$, the dynamics of the state probabilities is described by a symmetric rate matrix,
\begin{equation}
    \label{eq:stochastic-matr}
    {\bf Q}=
\begin{pmatrix}
-\mu_{\rm tot} & \mu_{01} & \mu_{10} & \mu_{11} \\
\mu_{01} & -\mu_{\rm tot} & \mu_{11} & \mu_{10} \\
\mu_{10} & \mu_{11} & -\mu_{\rm tot} & \mu_{01} \\
\mu_{11} & \mu_{10} & \mu_{01} & -\mu_{\rm tot}
\end{pmatrix},
\qquad \mu_{\rm tot}=\mu_{01}+\mu_{10}+\mu_{11}.
\end{equation}
The dynamics dramatically simplifies if we consider marginal fail and success complementary probabilities,
\[p_{\rm F,0}=p_{01}+p_{11},\quad p_{\rm S,0}=p_{00}+p_{10},\qquad  
  p_{\rm F,1}=p_{10}+p_{11},\quad p_{\rm S,1}=p_{00}+p_{01},
\]
as well as the fixed-parity complementary probabilities, 
\[p_{\rm odd}=p_{01}+p_{10},\quad p_{\rm even}=p_{00}+p_{11}.\]
The corresponding two-state stochastic matrices are also symmetrical, of the form $${\bf Q}^{(2)}(\mu)=\begin{pmatrix}-\mu&\mu\\ \mu&-\mu\end{pmatrix},$$
and have the expected marginal error rates for the two logical bits ($\mu_0$ and $\mu_1$) and for the third parity bit ($\mu_2$): 
\begin{equation}
    \label{eq:marginal-rates}
    \mu_0=\mu_{01}+\mu_{11},\quad \mu_1=\mu_{10}+\mu_{11},\text{ and }\mu_2=\mu_{10}+\mu_{01}.
\end{equation}  
In other words, the corresponding \emph{polarizations} (the probability differences) have simple exponential evolution, e.g., 
\[
S_{0}(\tau)\equiv p_{\rm S,0}(\tau)-p_{\rm F,0}(\tau)
=S_{0}(0)\,e^{-2\mu_0 \tau}.
\]
With the total probability constant, we get 
\begin{eqnarray}
    \label{eq:markov-p-0}
    p_{\rm F,0}(\tau)&=&{1\over2}\left[1-S_0(0)\,e^{-2\mu_0 \tau}\right],\\
    \label{eq:markov-p-1}
    p_{\rm F,1}(\tau)&=&{1\over2}\left[1-S_1(0)\,e^{-2\mu_{1}\tau}\right],\\
    \label{eq:markov-p-any}
    p_{\rm F,any}(\tau)&=&{3\over4}-{1\over4}\left[S_0(0)\,e^{-2\mu_0\tau}+S_1(0)\,e^{-2\mu_1 \tau}+S_2(0)\,e^{-2\mu_2 \tau}\right].
\end{eqnarray}
Evidently, each of the two marginal error probabilities for the logical observables relaxes as a single exponential, but not the net block error probability.  
These equations give the following expressions for the error probabilities over the duration of a single surface-code cycle:
\begin{equation}
\varepsilon_0=\frac{1-e^{-2\mu_{0}}}2,\qquad
\varepsilon_1=\frac{1-e^{-2\mu_{1}}}2,\qquad 
\varepsilon_L=\frac{3-e^{-2\mu_0}-e^{-2\mu_1}-e^{-2\mu_2}}4.
\label{eq:epsilon-01L}  
\end{equation}
Indeed, these expressions are just the above probabilities at $\tau=1$ (cycle duration in the chosen time units), conditioned on the absence of errors at the start of the cycle, which is equivalent to $S_i(0)=1$. 

Eqs.~\eqref{eq:markov-p-0}, \eqref{eq:markov-p-1}, and \eqref{eq:markov-p-any} are used:
\begin{itemize}
    \item To extract the net error rate $\mu_{\rm tot}$ for the numerical data in Sec.~\ref{sec:results}.  Here for each set of parameters we only had a single data point with $\ro=1$, i.e., the three probabilities $p_{\rm F,0}$, $p_{\rm F,1}$, and $p_{\rm F,any}$ at $\tau=3\rv+4$.  Thus, instead of the actual fitting, we set the initial polarizations $S_i(0)=1$, $i=0,1,2$, and solved for the rates $\mu_i$.
    \item To fit the experimental data in Sec.~\ref{sec:experimental}.  Here for each experiment we had 8 probability values for circuit variants with $1\le\ro\le8$; we fitted these data using six parameters, the three error rates and the initial polarizations.
    \item To obtain single-point error probability estimates in Fig.~\ref{fig:corr_markov_moving_average}.  Here, for each experiment and for each $\tau$ we extracted the rates from each data point (fixing the initial polarizations to the values from the full fit in item 2), and used these results to compute the single-cycle error probabilities  \eqref{eq:epsilon-01L}. 
\end{itemize}

\section{Numerical results with the uniform depolarizing noise model}
\label{app:uniform}

\begin{figure}[tbp]
  \centering
  \begin{minipage}[t]{0.4\linewidth}
    \centering
    \includegraphics[width=\linewidth]{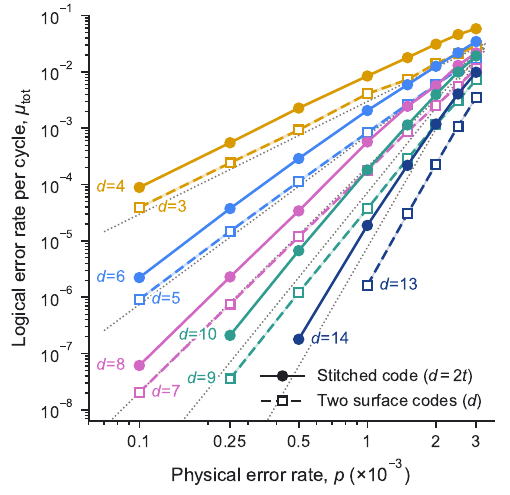}
    \caption{Same comparison as Fig.~\ref{fig:SI1000_p_vs_LER} under the uniform depolarizing model.}
    \label{fig:STD_p_vs_LER}
  \end{minipage}\hfill
  \begin{minipage}[t]{0.4\linewidth}
    \centering
    \includegraphics[width=\linewidth]{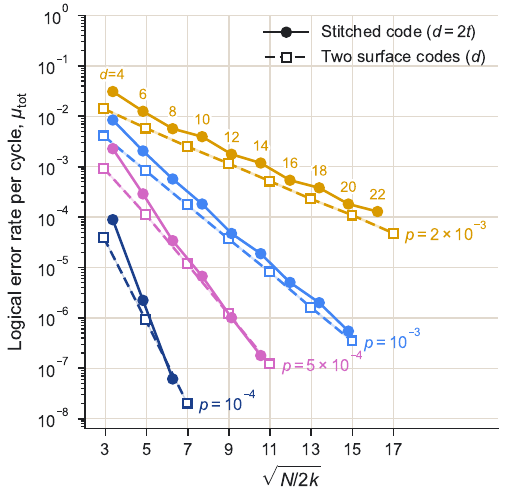}
    \caption{Same as Fig.~\ref{fig:footprint_SI1000} under the uniform depolarizing model; data of Fig.~\ref{fig:STD_p_vs_LER}.}
    \label{fig:footprint_std}
  \end{minipage}
\end{figure}

Figures~\ref{fig:STD_p_vs_LER} and~\ref{fig:footprint_std} repeat the comparisons of Figs.~\ref{fig:SI1000_p_vs_LER} and~\ref{fig:footprint_SI1000} under the uniform depolarizing model of Sec.~\ref{sec:results_setup}.
Under this model the stitched code is ahead from $N\approx 157$ at $p=10^{-4}$ and from $N\approx 445$ at $p=5\times10^{-4}$; at $p\geq 10^{-3}$ it requires more qubits, with the gap shrinking as $N$ grows.

\section{Detector-level diagnostics}
\label{app:first_appendix}

This appendix collects diagnostics of the hardware analysis: finite-duration single-point checks of the per-cycle rates, a comparison of detector statistics with a noisy circuit-level simulation, and the decoding role of the outer-code detectors.

\subsection{Detector probabilities and correlations}
\label{app:detector_diagnostics}

\begin{figure}[tbp]
    \centering
    \includegraphics[width=0.48\linewidth]{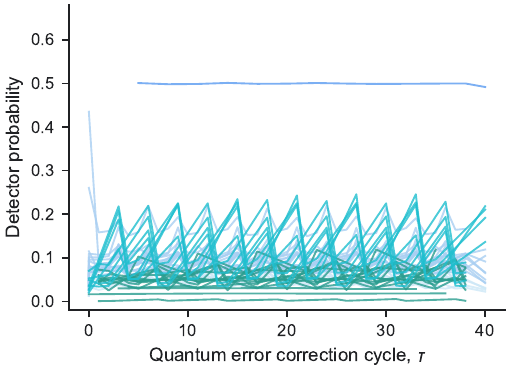}
    \hfill
    \includegraphics[width=0.48\linewidth]{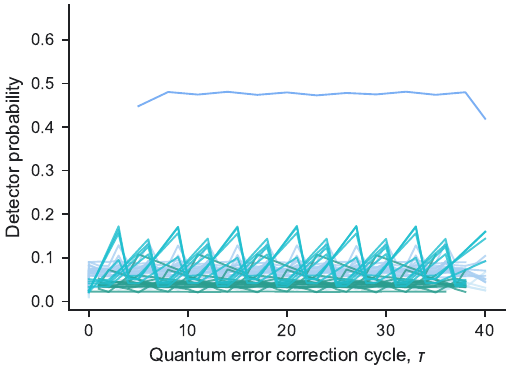}
    \caption{Left: Grouped detection-event fractions for the $d=6$, $\ro=6$ $ZZ$ hardware experiment.  Right: same, but from a simulation of the same circuit under the distributed noise model of Appendix~\ref{app:detector_diagnostics} with $200{,}000$ shots. The high-fire-fraction traces are from the outer-code detectors, whose physical weight $4t=12$ makes them sensitive to many localized fault mechanisms.}
    \label{fig:det_frac}
\end{figure}

\begin{figure}[tbp]
    \centering
    \includegraphics[width=0.45\linewidth]{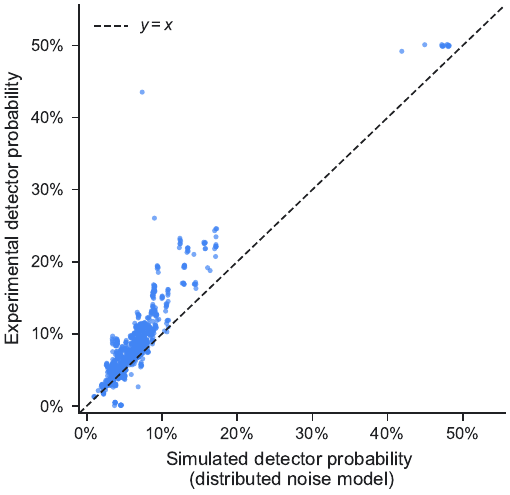}
    \hfill 
    \includegraphics[width=0.45\linewidth]{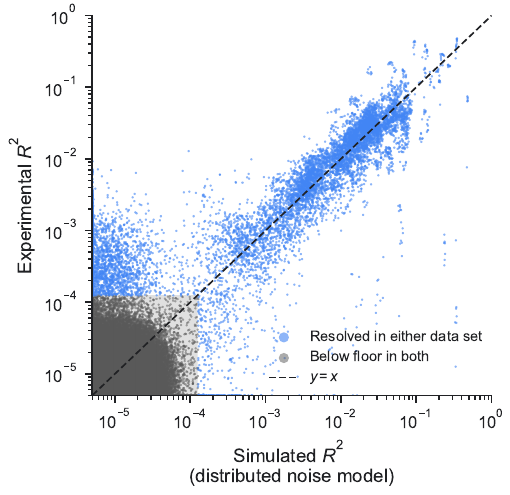}
    \caption{Measured detector statistics vs.\ the simulated values. Left: Single-detector fire probabilities  from Fig.~\ref{fig:det_frac}. Right: Squared Pearson correlation $R^2$ of every detector pair; points in the gray region, with both values $R^2<25/N_{\rm shot}$, do not have sufficient statistics to be relevant for comparison.}
    \label{fig:appendix_detector_analysis}
\end{figure}

\begin{figure}[tbp]
    \centering
    \includegraphics[width=0.48\linewidth]{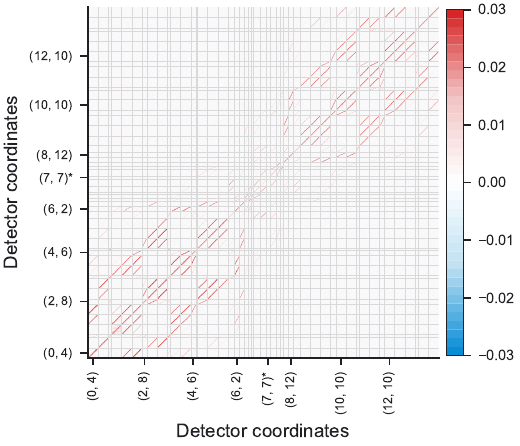}
    \hfill 
    \includegraphics[width=0.48\linewidth]{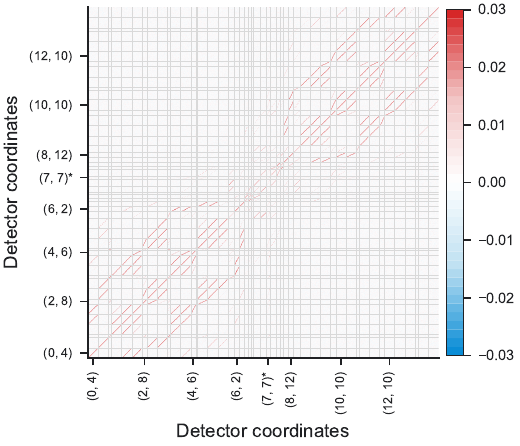}
    \caption{Pearson correlation matrices of all detectors for the measured data (left) and the simulated values (right). Detectors are grouped by site $(x,y)$ and ordered by time within each site; the outer-code detectors at the corridor intersection form their own block located near the center (coordinates labeled with ``*''). Diagonal elements are set to zero.}
    \label{fig:appendix_correlation_matrices}
\end{figure}

Figures~\ref{fig:det_frac}--\ref{fig:appendix_correlation_matrices} compare the $d=6$, $\ro=6$ $ZZ$ hardware data with a simulation of the same executed circuit, including its single-qubit echo pulses, each with $N_{\rm shot}=200{,}000$ shots.
The simulation uses a distributed noise model built from the per-qubit and per-pair calibration of the device (Fig.~\ref{fig:hardware_err}): each single-qubit gate and CZ gate is followed by depolarizing noise with its measured Pauli error, each measurement outcome is flipped with the measured readout error of its qubit, and each idle qubit receives Pauli-twirled amplitude and phase damping set by its $T_1$ and $T_\phi$ over the duration of the layer.
Layer durations follow Ref.~\cite{Google2025BelowThreshold}.
The simulation reproduces the grouped structure of the detection-event fractions, including the saturated outer-code detectors and the periodic peaks on the corridor-adjacent patch detectors [Fig.~\ref{fig:det_frac}].
Across all $1{,}384$ detectors, the hardware and simulated fire probabilities have a Pearson correlation of $0.93$, with means of $8.0\%$ and $6.3\%$, respectively [Fig.~\ref{fig:appendix_detector_analysis}, left].
For detector pairs, we treat a squared Pearson correlation $R^2$ as resolved when it exceeds $25/N_{\rm shot}$, that is, when the correlation exceeds five standard deviations of its shot noise.
Of the $11{,}455$ pairs resolved in either data set, $6{,}505$ are resolved in both, and their $\log R^2$ values have a correlation of $0.89$ [Fig.~\ref{fig:appendix_detector_analysis}, right].
Of the others, $4{,}522$ are resolved only in the hardware data and $428$ only in the simulation; together with the lower simulated mean fire probability, this indicates hardware noise, such as leakage and crosstalk, beyond the calibrated gate, readout, and idle errors.
The full correlation matrices in Fig.~\ref{fig:appendix_correlation_matrices} show the same local structure in both data sets.

\subsection{Decoding role of the outer-code detectors}
\label{app:outer_role}

At $d=6$ every outer-code detector fires with probability within about $0.01$ of $1/2$ [Fig.~\ref{fig:det_frac}, left], so in isolation it cannot be distinguished from a random bit.
Its value to the decoder is conditional on the rest of the syndrome.
To measure it, we decode every shot twice with correlated matching: once with the measured value of a given outer-code detector and once with that value flipped, keeping all other detectors fixed.
We call a shot decisive for the detector when the two decoded logicals differ.

For the $d=6$, $\ro=1$ $ZZ$ data, each $\overline{ZZZZ}$ outer-code detector is decisive in $81\%$ of the shots.
In these shots the measured value gives the correct logical outcome in $82.7\%$ of cases and the flipped value in $4.7\%$, so whenever exactly one of the two is correct, the measured value is the correct one $95\%$ of the time.
The $XX$ data give $78\%$ decisive shots and $93\%$ for the same ratio.
Replacing the outer-code detectors by independent random bits with the same fire probability leaves the decisive fraction near $80\%$ but makes the two outcomes equally likely to be correct.
The $\overline{XXXX}$ detectors of a $ZZ$ experiment, and vice versa, are decisive in at most $11\%$ of the shots, since they do not act on the measured observables.

The decoding value of the outer-code detectors is seen directly by decoding the same shots with the same correlated-matching decoder but without them (their detection events are ignored and the corresponding error mechanisms become unmatched). At $d=6$ the logical error rate per cycle, from the Markov fit of Sec.~\ref{sec:extended_cycles}, rises from $3.80\%$ to $4.65\%$ ($ZZ$) and from $4.10\%$ to $5.19\%$ ($XX$); at $d=4$ the block error probability after a single outer-code period rises from $0.19$ ($ZZ$) and $0.22$ ($XX$) to $0.48$, close to its saturation value of $3/4$. Replacing the outer-code detectors by independent random bits is worse still, since the decoder then matches spurious events.

\Needspace*{16\baselineskip}
\section{Stability experiment}
\label{app:stability}

\begin{figure}[h!]
  \centering
  \includegraphics[width=0.32\linewidth]{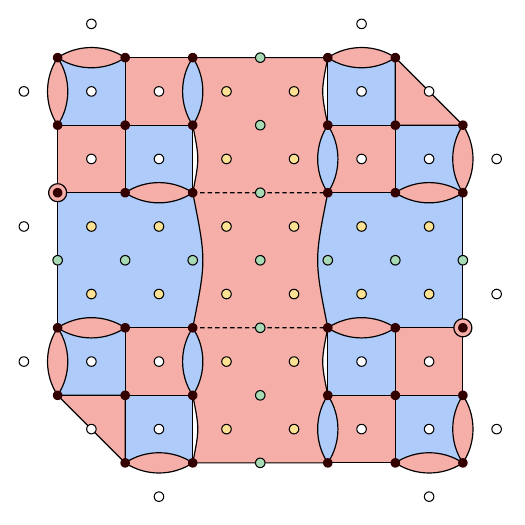}
  \hfill
  \includegraphics[width=0.66\linewidth]{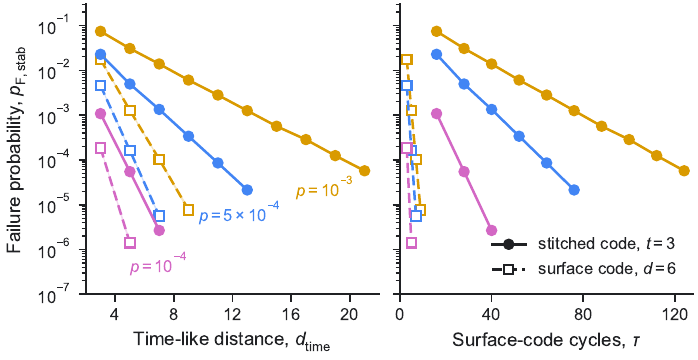}
  \caption{Left: stitched stability circuit for $t=3$ (colors as in Fig.~\ref{fig:layout_d3}): patch generators and $G_X$ (vertical band) as Stim detector slices~\cite{Gidney2021Stim}, and $G_Z$ (horizontal band, under $G_X$) on the data qubits that the $\overline{Z}$-corridor couples to; circles: single-qubit generators of the two corner qubits on these rows. Middle and right: stability failure probability $p_{\rm F,stab}$ of the stitched code ($t=3$, $\rv=2$; filled circles) and the surface code of distance $6$ (open squares) versus $d_{\rm time}$ and versus $\tau$, at the $p$ written next to the curves; SI1000 noise, Corr-MWPM decoding; bands: 95\% Wilson intervals.}
  \label{fig:stability}
\end{figure}

\begin{wraptable}{r}{0.45\linewidth}
\centering
\small
\vspace{-1.1em}
\begin{tabular}{@{}l c c c c c@{}}
\toprule
$\ro$ & \makebox[2.2em]{$1$} & \makebox[2.2em]{$2$} & \makebox[2.2em]{$3$} & \makebox[2.2em]{$4$} & \makebox[2.2em]{$5$} \\
\midrule
$d_{\rm time}$ & $2$ & $3$ & $4$ & $5$ & $6$ \\
\addlinespace[4pt]
& \multicolumn{5}{c}{surface-code cycles $\tau$} \\
\cmidrule(l){2-6}
$\rv=1$ & $7$ & $11$ & $15$ & $19$ & $23$ \\
$\rv=2$ & $10$ & $16$ & $22$ & $28$ & $34$ \\
$\rv=3$ & $13$ & $21$ & $29$ & $37$ & $45$ \\
\bottomrule
\end{tabular}
\caption{Time-like distance $d_{\rm time}$ of the stitched code for $t=3$ (exact, dist-m4ri~\cite{distm4ri}; the same for $\rv=1$, $2$, $3$) and circuit duration $\tau$; for the surface code, $d_{\rm time}=\tau$~\cite{Gidney2022Stability}.}
\label{tab:stability_distance}
\vspace{-1.4em}
\end{wraptable}
A memory experiment tests the ability to move logical observables through time; a stability experiment~\cite{Gidney2022Stability} tests the ability to move them through space, that is, to determine the value of a product of stabilizer generators, as needed, for example, by measurement-based logical operations such as lattice surgery~\cite{Horsman2012,FowlerGidney2018,Litinski2019}.
Following Ref.~\cite{Gidney2022Stability}, we construct a stability experiment for the outer-code generator $G_X$ of the stitched code (Fig.~\ref{fig:stability}, left): the patches have $X$-type boundaries on the three sides that do not face the $\overline{X}$-corridor, so that their $X$-type generators multiply to $G_X$.
To keep the circuit faithful, $G_Z$ and the $Z$-type generators are still measured, although the observable does not depend on them.
Since $G_Z$ may flip four $X$-type generators, one per patch, by a common random sign, we track these generators in pairs at the two ends of the $\overline{Z}$-corridor, so that every fault mechanism decomposes into graphlike components.
The time-like distance is then the number of $G_X$ outcomes, $d_{\rm time}=\ro+1$ (Table~\ref{tab:stability_distance}).

We compare it with the surface-code stability circuit of Ref.~\cite{Gidney2022Stability} at distance $2t$, using SI1000 noise~\cite{GidneyNewmanFowlerBroughton2021} and Corr-MWPM decoding for both.
For $t=3$, each outer-code round lowers the failure probability $p_{\rm F,stab}$ by a factor of $1.5$, $2.0$, and $4.5$ at $p=10^{-3}$, $5\times10^{-4}$, and $10^{-4}$ (Fig.~\ref{fig:stability}, middle): the outer code is fault tolerant in time.
This protection comes at a cost in time, since one unit of $d_{\rm time}$ takes $2(\rv+1)$ surface-code cycles instead of one.
At $p=10^{-3}$, reaching $p_{\rm F,stab}=10^{-3}$ takes about $80$ cycles, compared with about $5$ for the surface code (Fig.~\ref{fig:stability}, right).
Since the optimal $\rv$ depends on $t$ and $p$ (Sec.~\ref{sec:results_rv}), the number of cycles needed to reach a given $p_{\rm F,stab}$ requires a careful analysis in each case, but in all cases we studied ($t=3$, $5$) it is much larger than for the surface code.

Thus GHZ corridors serve memory well, since only changes of the outer-code generators matter there, but any logical operation that relies on corridor outcomes inherits their slow accumulation of time-like distance, and we do not expect GHZ corridors to be a competitive replacement for lattice surgery in computation.
Logical operations on stitched codes can instead avoid relying on corridor outcomes, for example through lattice surgery on the patch boundaries that do not face the corridors, or through code automorphisms~\cite{GrasslRoetteler2013,BreuckmannBurton2024,Sayginel2025}, such as the physical-qubit permutations of the $[\,\![4,2,2]\,\!]$ code (Sec.~\ref{sec:prelim_422}).
We leave logical operations on stitched codes, and stitched codes with other small outer codes, to future work.

\end{document}